\documentclass[smallextended]{svjour3}       
\smartqed  
\usepackage[utf8]{inputenc}
\usepackage[T1]{fontenc}
\usepackage{amsmath,amssymb,amsfonts}
\usepackage{algorithmic}
\usepackage{graphicx,subfigure}
\usepackage{textcomp}
\usepackage{booktabs}
\usepackage{framed}
\usepackage{float}
\usepackage{xcolor}
\usepackage{footnote}
\usepackage{xfrac}
\usepackage{balance}
\usepackage{hyperref}
\usepackage{array}
\usepackage{subfigure}
\usepackage{changepage}
\usepackage{listings}

\newcolumntype{L}[1]{>{\raggedright\let\newline\\\arraybackslash\hspace{0pt}}m{#1}}

\definecolor{mygreen}{rgb}{0,0.6,0}
\definecolor{lightgreen}{rgb}{0.6,0.9,0.6}
\definecolor{lightyellow}{rgb}{0.9,0.9,0.6}
\definecolor{lightorange}{rgb}{0.9,0.8,0.6}
\definecolor{lightred}{rgb}{0.9,0.7,0.7}
\definecolor{mygray}{rgb}{0.5,0.5,0.5}
\definecolor{lightgray}{rgb}{0.8,0.8,0.8}
\definecolor{mymauve}{rgb}{0.58,0,0.82}

\usepackage{pifont}

\lstdefinestyle{CStyle} {
    language=C,
    backgroundcolor=\color{white},   
    basicstyle=\ttfamily\scriptsize,        
    breaklines=true,                 
    captionpos=b,                    
    commentstyle=\color{mygray}\bfseries,    
    escapeinside={\%*}{*)},          
    keywordstyle=\color{blue},       
    stringstyle=\color{mymauve},     
    frame=single,
    numbers=left,
    stepnumber=1,
    xleftmargin=2em,
    escapeinside={/*!}{!*/},
    moredelim=**[l][\color{mygreen}]{+\ },
    moredelim=*[l][\color{red}]{-\ }
}

\lstdefinestyle{JavaStyle} {
    language=Java,
    backgroundcolor=\color{white},   
    basicstyle=\ttfamily\scriptsize,        
    breaklines=true,                 
    captionpos=b,                    
    commentstyle=\color{mygray}\bfseries,    
    escapeinside={\%*}{*)},          
    keywordstyle=\color{blue},       
    stringstyle=\color{mymauve},     
    frame=single,
    numbers=left,
    stepnumber=1,
    xleftmargin=2em,
    escapeinside={/*!}{!*/},
    moredelim=**[l][\color{mygreen}]{+\ },
    moredelim=*[l][\color{red}]{-\ }
}

\lstdefinestyle{PHPStyle} {
    language=PHP,
    alsolanguage=HTML,
    backgroundcolor=\color{white},   
    basicstyle=\ttfamily\scriptsize,        
    breaklines=true,                 
    captionpos=b,                    
    commentstyle=\color{mygray}\bfseries,    
    escapeinside={\%*}{*)},          
    keywordstyle=\color{blue},       
    stringstyle=\color{mymauve},     
    frame=single,
	numbers=left,
    stepnumber=1,
	xleftmargin=2em,
    escapeinside={/*!}{!*/},
    moredelim=**[l][\color{mygreen}]{+\ },
    moredelim=*[l][\color{red}]{-\ }
}

\newcounter{lstannotation}
\renewcommand{\thelstannotation}{\ding{\number\numexpr181+\arabic{lstannotation}}}
\newcommand{\annotation}[1]{\refstepcounter{lstannotation}\label{#1}\thelstannotation}

\newboolean{showcomments}
\setboolean{showcomments}{true}

\ifthenelse{\boolean{showcomments}}
  {\newcommand{\nb}[3]{
  {\color{#2}\small\fbox{\bfseries\sffamily\scriptsize#1}}
  {\color{#2}\sffamily\small$\triangleright~$\textit{\small #3}$~\triangleleft$}
  }
  }
  {\newcommand{\nb}[3]{}
  }

\makeatletter
\newcommand\footnoteref[1]{\protected@xdef\@thefnmark{\ref{#1}}\@footnotemark}
\makeatother

\begin{document}

\title{Fixing Vulnerabilities Potentially Hinders Maintainability}

\author{Sofia Reis         \and
        Rui Abreu           \and
        Luis Cruz
}

\institute{Sofia Reis \at
			INESC ID and IST, University of Lisbon, Lisbon, Portugal\\
            \email{sofia.o.reis@tecnico.ulisboa.pt}   
			\and
			Rui Abreu \at
			INESC ID and FEUP, University of Porto, Porto, Portugal\\
			\email{rui@computer.org}
           \and
			Luis Cruz \at
			Delft University of Technology, Delft, The Netherlands\\
			\email{L.Cruz@tudelft.nl}
}

\date{Received: date / Accepted: date}

\maketitle

\begin{abstract}
Security is a requirement of utmost importance to produce 
high-quality software. However, there is still a considerable amount 
of vulnerabilities being discovered and fixed almost weekly. We 
hypothesize that developers affect the maintainability of their 
codebases when patching vulnerabilities. This paper evaluates the 
impact of patches to improve security on the maintainability of 
open-source software. Maintainability is measured based on the 
Better Code Hub’s model of 10 guidelines on a dataset, including 
1300 security-related commits. Results show evidence of a trade-off 
between security and maintainability for $41.90\%$ of the cases, i.e., developers 
may hinder software maintainability. Our analysis shows that $38.29\%$ of patches increased
software complexity and $37.87\%$ of patches increased the percentage 
of LOCs per unit. The implications of our study 
are that changes to codebases while patching vulnerabilities need to 
be performed with extra care; tools for patch risk assessment should 
be integrate into the CI/CD pipeline; computer science curricula 
needs to be updated; and, more secure programming languages are 
necessary.
\end{abstract}

\section{Introduction}
Software quality is important because it is ultimately related to 
the overall cost of developing and maintaining software 
applications, security and safety~\cite{slaughter1998evaluating}. Software quality 
characteristics include, but are not limited to functional 
correctness, reliability, usability, maintainability, evolvability 
and security. Security is an essential non-functional requirement 
during the development of software systems. In $2011$, the 
International Organization for Standardization (ISO) issued an 
update for software product quality ISO/IEC 25010 considering 
\emph{Security} as one of the main software product quality 
characteristics~\cite{iso:2011}. However, there is still 
a considerable amount of vulnerabilities being discovered and fixed, 
almost weekly, as disclosed by the Zero Day Initiative 
website\footnote{Zero Day Initiative website available at 
\url{https://www.zerodayinitiative.com/advisories/published/} 
(Accessed on \today{})}. 

Researchers found a correlation between the presence of 
vulnerabilities on software and code complexity~\cite{shin2010evaluating,10.1145/1774088.1774504}. 
Security experts claim that complexity hides bugs that may result in 
security vulnerabilities~\cite{mcgraw2004software,schneier2006beyond}. In 
practice, an attacker only needs to find one way into the system 
while a defender needs to find and mitigate all the security issues. 
Complex code is difficult to understand, maintain and 
test~\cite{1702388}. Thus, the task of a developer gets more challenging as the 
codebase grows in size and complexity. But the risk can be minimized 
by writing clean and maintainable code. 

ISO describes \textit{software maintainability} as ``the degree of 
effectiveness and efficiency with which a software product or system 
can be modified to improve it, correct it or adapt it to changes in 
environment, and in requirements'' on software quality ISO/IEC 
25010~\cite{iso:2011}. Thereby, maintainable security may be 
defined, briefly, as the degree of effectiveness and efficiency with 
which software can be changed to mitigate a security 
vulnerability---corrective maintenance.
However, many developers still lack knowledge on the best 
practices to deliver and maintain secure and high-quality 
software~\cite{Pothamsetty:2005:SEL:1107622.1107635,8077802}. In a 
world where zero-day vulnerabilities are constantly emerging, 
mitigation needs to be fast and efficient. Therefore, it 
is important to write maintainable code to support the production of 
more secure software---maintainable code is less complex and, 
consequently, less prone to
vulnerabilities~\cite{shin2010evaluating,10.1145/1774088.1774504}---
and, prevent the introduction of new vulnerabilities. 

As ISO does not provide any specific guidelines/formulas to 
calculate maintainability, we resort to Software Improvement Group 
(SIG\footnote{SIG's website: 
\url{https://www.sig.eu/} (Accessed on \today{})})'s web-based source 
code analysis service Better Code Hub (BCH)\footnote{BCH's 
website: \url{https://bettercodehub.com/} (Accessed on \today{})}  
to compute the software compliance with a set of $10$ 
guidelines/metrics to produce quality software based on ISO/IEC 
$25010$~\cite{Visser:2016:OREILLY}. SIG 
has been helping business and technology leaders drive their organizational 
objectives by fundamentally improving the health and security of 
their software applications for more than 20 years. Their 
models are scientifically proven and certified~\cite{4335232,5609747,6113040,baggen2012}.

There are other well-known 
standards and models that have been proposed to increase software 
security: Common Criteria~\cite{common:2009} which received
negative criticism regarding the costs associated and poor technical 
evaluation; the OWASP Application Security Verification 
Standard (ASVS)~\cite{oswap:2009} which is focused only on web 
applications, and a model proposed by Xu et al. ($2013$)
~\cite{6616351} for rating software security (arguably, it was one 
of the first steps taken by SIG to introduce security on their 
maintainability model). Nevertheless, our study uses BCH to provide 
an assessment of maintainability in software for the following 
reasons: BCH integrates a total of $10$ different code metrics; and, 
code metrics were empirically validated in previous 
work~\cite{Bijlsma:2012:FIR:2317098.2317124,8530041,8919169,8785997}.

Static analysis tools (SATs) have been built to detect software 
vulnerabilities automatically (e.g., FindBugs, Infer, and more). Developers
use those tools to locate the issues in the code. However,
while performing the patches to those issues, SATs cannot provide 
information on the quality of the patch. 
Improving software security is not a trivial task and requires 
implementing patches that might affect software maintainability. 
We hypothesize that some of these patches may 
have a negative impact on the software maintainability and, 
possibly, even be the cause of the introduction of new 
vulnerabilities---harming software reliability and introducing 
technical debt. Research found that $34\%$ of the security patches 
performed introduce new problems and $52\%$ are incomplete and do not 
fully secure systems~\cite{10.1145/3133956.3134072}. Therefore, in this paper, 
we present an empirical study on the impact of patches of 
vulnerabilities on software maintenance across open-source software.
We argue that tools that assess these type of code metrics may
complement SATs with valuable information to help the developer
understand the risk of its patch.

From a methodological perspective, we leveraged a dataset of $1300$ 
security patches collected from open-source software. We calculate 
software maintainability before and after the patch to measure its 
impact. This empirical study presents evidence that changes applied 
in the codebases to patch vulnerabilities affect code 
maintainability. Results also suggest that developers 
should pay different levels of attention to different severity 
levels and classes of weaknesses when patching vulnerabilities. We 
also show that patches in programming languages such as, 
\emph{C/C++}, \emph{Ruby} and \emph{PHP}, may have a more negative 
impact on software maintainability. Little information is known 
about the impact of security patches on software 
maintainability. Developers need to be aware of the impact of their 
changes on software maintainability while patching security 
vulnerabilities. The harm of maintainability can increase the time 
of response of future mitigations or even of other maintainability 
tasks. Thus, it is of utmost importance to find means to assist 
mitigation and reduce its risks. With this study, we intend to 
highlight the need for tools to assess the impact of patches on 
software maintainability~\cite{4724577}; the importance of 
integrating maintainable security in computer science curricula; 
and, the demand for better programming languages, designed by 
leveraging security principles~\cite{kurilova2014wyvern,10.1145/2489828.2489830}. 
 
This research performs the following main contributions:
\begin{itemize}
  \item Evidence that supports the trade-off between security and 
  maintainability: developers may be hindering software 
  maintainability while patching vulnerabilities.
	\item An empirical study on the impact of security patches on 
	software maintainability (per guideline, severity, weakness and 
	programming language).
	\item A replication package with the scripts and data created to 
	perform the empirical evaluation for reproducibility. Available 
	online: \url{https://github.com/TQRG/maintainable-security}.
\end{itemize}
This paper is structured as follows: Section~\ref{sec:motivation} 
introduces an example of a security patch of a known vulnerability 
found in the protocol implementation of 
OpenSSL\footnote{\label{openssl}OpenSSL is a toolkit that
contains open-source implementations of the SSL and TLS cryptographic
protocols. Repository available at 
\url{https://github.com/openssl/openssl} (Accessed on \today{})}; 
Section~\ref{sec:methodology} describes the methodology used to 
answer the research questions; Section~\ref{sec:results} presents 
the results and discusses their implications; 
Section~\ref{sec:implications} elaborates on the implications
developers should consider in the future; Section~\ref{sec:threats} 
enumerates the threats to the validity of this study; 
Section~\ref{sec:rw} describes the different work and existing
literature in the field of study; and, finally, 
Section~\ref{sec:conclusions} concludes the main findings and 
elaborates on future work.

\section{Motivation and Research Questions}\label{sec:motivation}

As an example, consider the patch of the TLS state machine protocol 
implementation in OpenSSL\footnoteref{openssl} to address a memory 
leak flaw found in the way how OpenSSL handled TLS status request 
extension data during session renegotiation, and where a malicious 
client could cause a Denial-of-Service (DoS) attack via large Online 
Certificate Status Protocol (OCSP) Status Request extensions when 
OCSP stapling support was enabled. OCSP stapling, formally known 
as the TLS Certificate Status Request extension, is a standard for 
checking the revocation status of certificates. 

This vulnerability is listed at the Common Vulnerabilities and 
Exposures dictionary as CVE-$2016$-$6304$~\footnote{CVE-$2016$-$6304$
details available at 
\url{http://cve.mitre.org/cgi-bin/cvename.cgi?name=CVE-2016-6304}
(Accessed on \today{})}. It is amongst the vulnerabilities studied 
in our research. The snippet, in Listing~\ref{lst:vuln}, presents 
the changes performed on the
\emph{ssl/t1$\_$lib.c} file\footnote{CVE-$2016$-$6304$ fix available 
 at
\url{https://github.com/openssl/openssl/commit/e408c09bbf7c3057bda4b8d20bec1b3a7771c15b}
(Accessed on \today{})} by the OpenSSL developers to patch the 
vulnerability. Every SSL/TLS connection begins with a handshake 
which is responsible for the negotiation between the two parties. 
The OSCP Status Request extension allows the client to verify the 
server certificate and enables a TLS server to include its response 
in the handshake. The problem in CVE-$2016$-$6304$ is a flaw in the 
logic of OpenSSL that does not handle memory efficiently when large 
OCSP Status Request extensions are sent each time a client requests 
renegotiation. This was possible because the OCSP responses IDs 
were not released between handshakes. Instead, they would be 
allocated again and again. Thus, if a malicious client does it 
several times it may lead to an unbounded memory growth on the 
server and, eventually, lead to a DoS attack through memory 
exhaustion. 

The code changes performed to patch the CVE-$2016$-$6304$ vulnerability are 
presented in Listing~\ref{lst:vuln}. The \texttt{sk\_OCSP\_RESPID\_pop\_free} function 
(\ref{lst:func1}) removes any memory allocated to the OCSP response IDs 
(\texttt{OCSP\_RESPID}s) from a previous handshake to prevent unbounded 
memory growth---which was not being performed before.
After releasing the unbounded memory, the logic condition in 
\ref{lst:func3} was shifted to \ref{lst:func2} which is responsible 
for handling the application when no OCSP response IDS are 
allocated. After the patch, in the new version of the software, the 
condition is checked before the package processing 
instead of after. Thereby, the system avoids the increase of 
unbounded memory (and a potential DoS attack).

Patching this vulnerability seems a rudimentary task. Yet,
a considerable amount of changes were performed in the codebase 
which yielded a negative impact on software maintainability.
While patching, the developer introduced $6$ new lines in a method 
already with a large number of lines of code and introduced more 
complexity to the code with $2$ new branch points, which disrupt two of 
the guidelines proposed by the Software Improvement Group 
(SIG) for building maintainable software~\cite{Visser:2016:OREILLY}: 
\emph{Write Short Units of Code} and \emph{Write Simple Units of 
Code}. 

\medskip
\setcounter{lstannotation}{0}
\begin{lstlisting}[style={CStyle}, caption={Patch provided by OpenSSL developers to the
CVE-2016-6304 vulnerability on file ssl/t1\_lib.c},label={lst:vuln}]
static int ssl_scan_clienthello_tlsext(SSL *s, PACKET *pkt, int *al){ 
 // [snip]
+  sk_OCSP_RESPID_pop_free(s->tlsext_ocsp_ids, OCSP_RESPID_free); /*!\annotation{lst:func1}!*/
+  if (PACKET_remaining(&responder_id_list) > 0) { 
+     s->tlsext_ocsp_ids = sk_OCSP_RESPID_new_null();
+     if (s->tlsext_ocsp_ids == NULL) { /*!\annotation{lst:func2}!*/
+        *al = SSL_AD_INTERNAL_ERROR;
+        return 0;
+     }
+  } else {
+     s->tlsext_ocsp_ids = NULL;
+  }

   while (PACKET_remaining(&responder_id_list) > 0) {
     OCSP_RESPID *id;
     PACKET responder_id;
     const unsigned char *id_data;
     if (!PACKET_get_length_prefixed_2(&responder_id_list, &responder_id) || PACKET_remaining(&responder_id) == 0) {
          return 0;
     }

-  if (s->tlsext_ocsp_ids == NULL 
-      && (s->tlsext_ocsp_ids = 
-      sk_OCSP_RESPID_new_null()) == NULL) { /*!\annotation{lst:func3}!*/
-    *al = SSL_AD_INTERNAL_ERROR;
-    return 0;
-  }

 // [snip]
 }
\end{lstlisting}

Software maintainability is designated as the degree to which an 
application is understood, repaired, or enhanced. In this paper, our 
concern is to study whether while improving software security, 
developers are also hindering software maintainability. This is 
important because software 
maintainability is approximately $75\%$ of the cost related to a 
project. To answer the following three research questions, we use 
two datasets of security 
patches~\cite{Reis:2017:IJSSE,10.1109/MSR.2019.00064} to measure the 
impact of security patches on the maintainability of open-source 
software. 

\textit{\textbf{RQ1: What is the impact of security patches on the
maintainability of open-source software?}} Often, security flaws 
require patching code to make software more secure. However, 
\textbf{there is no evidence yet of how security patches impact the
maintainability of open-source software}. We hypothesize that 
developers tend to introduce technical debt in their software when 
patching software vulnerabilities because they tend not to pay enough 
attention to the quality of those patches. To address it, 
we follow the same methodology as previous research~\cite{8919169} and compute 
the maintainability of $1300$ patches using the \emph{Better Code Hub} tool. 
We present the maintainability impact by guideline/metric, overall score, 
severity, and programming language.

\textit{\textbf{RQ2: Which weaknesses are more likely to
affect open-source software maintainability?}} There are security 
flaws that are more difficult to patch than others. For instance, 
implementing secure authentication is not as easy as patching a
cross-site scripting vulnerability since the latter can be fixed
without adding new lines of code/complexity to the code. A typical 
fix for the cross-site scripting vulnerability is presented in 
Listing~\ref{lst:fix}. The developer added the function 
\texttt{htmlentities} to escape the data given by the variable
\texttt{\$\_['file']}. We hypothesize that security patches for 
different weaknesses can have different impacts on software maintainability. 
\textbf{Understanding which weaknesses are more likely to increase maintainability 
issues is one step toward bringing awareness to security engineers of what 
weaknesses need more attention.} The taxonomy of security patterns used to answer this 
question is the one provided by the Common Weakness Enumeration
(CWE). \emph{Weakness}, according to the Common Weakness Enumeration 
(CWE) glossary, is a type of code-flaw that could contribute to the 
introduction of vulnerabilities within that product. In this study, maintainability 
is measured separately for each weakness.
\setcounter{lstannotation}{0}
\begin{lstlisting}[style={PHPStyle}, caption={Fix provided by \texttt{nextcloud/server} 
  developers to a Cross-Site Scripting vulnerability},label={lst:fix}]
   <p class='hint'>
    <?php
-   if(isset($_['file'])) echo $_['file']
+   if(isset($_['file'])) echo htmlentities($_['file'])
    ?>
   </p>
\end{lstlisting}
\textit{\textbf{RQ3: What is the impact of security patches versus 
regular changes on the maintainability of open-source software?}}
Performing a regular change/refactoring, for instance, to improve the name of 
a variable or function is different than performing a security patch.
Therefore, we also computed the maintainability of random regular commits using 
the \emph{Better Code Hub} tool---baseline. We use them to 
understand \textbf{how maintainability evolves when security patches 
are performed versus when they are not}.
\section{Methodology}\label{sec:methodology}

\begin{figure}[h]
	\centering 	
    \begin{adjustwidth}{-0.5cm}{-0.5cm}  
      \vspace{-2.5cm} 
	\includegraphics[width=1.1\textwidth]{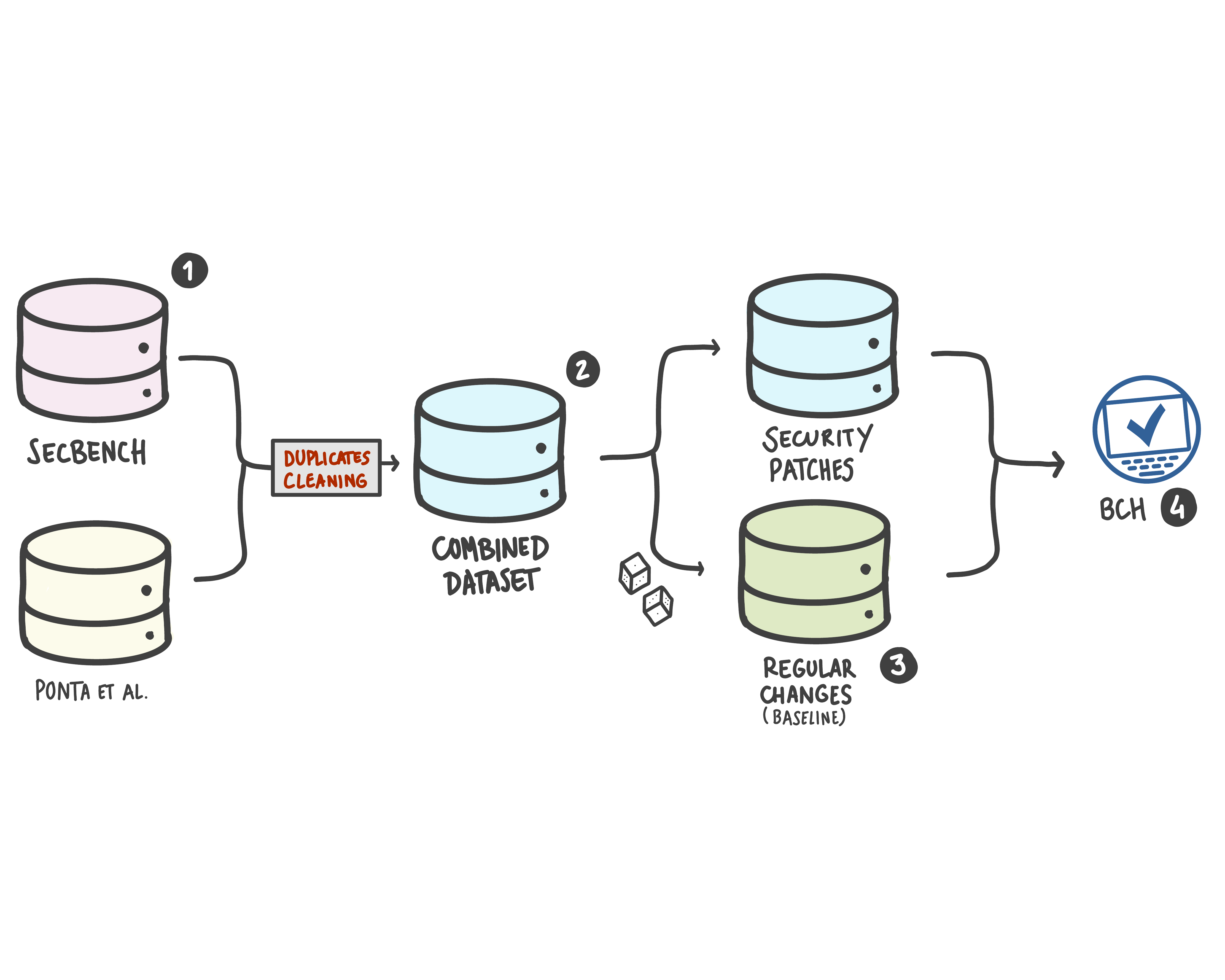}
  \vspace{-3cm} 
  \caption{Study Methodology}
	\label{fig:met}
	 \end{adjustwidth}
\end{figure}

In this section, we discuss the methodology used to measure
the impact of security patches on the maintainability of open-source
software. The methodology comprises the following steps, as 
illustrated in Figure~\ref{fig:met}.
\begin{enumerate}
	\item Combine the datasets from related work that classify
	the activities of developers addressing security-oriented 
  patches~\cite{reis2017secbench,10.1109/MSR.2019.00064}. The
  duplicated patches were tossed.
	\item
	Extract relevant data (e.g., owner and name of
	the repository, sha key of the vulnerable version, sha key of 	
	the fixed version) from the combined dataset 	
	containing $1300$ security patches collected from open-source 	
	software available on GitHub.
  \item Two baselines of regular changes were collected:
  \textit{random-baseline} (for each security commit, a random change 
  was collected from the same project) and \textit{size-baseline} (for 
  each security commit, a random change with the same size
  was collected from the same project). Our goal is to evaluate the impact 
	of regular changes on the maintainability of open-source 
	software.
  \item Use the Software Improvement Group (SIG)'s web-based source 
  code analysis service \emph{Better Code Hub} (BCH)
  to quantify maintainability for both security and regular commits. 
  BCH evaluates the codebase available in the default branch of a GitHub project. 
  We created a tool that pulls the codebase of each commit of our dataset 
  to a new branch; it sets the new branch as the default branch; and, runs 
  the BCH analysis on the codebase; after the analysis is finished, the tool saves 
  the BCH metrics results to a cache file.
\end{enumerate}
\subsection{Datasets}

We use a combined dataset of $1300$ security patches which is 
the outcome of mining and manually inspecting a total of $312$ 
GitHub projects. The combined dataset integrates two different
works: Secbench~\cite{reis2017secbench,Reis:2017:IJSSE} and Ponta et al.~\cite{10.1109/MSR.2019.00064}.

\textbf{Reis and Abreu 
($2017$)} mined open-source software aiming at the extraction of 
real---created by developers---patches of security vulnerabilities 
to test and assess the performance of static analysis 
tools~\cite{reis2017secbench,Reis:2017:IJSSE} since using hand-seeded test cases or 
mutations can lead to misleading assessments of the capabilities of 
the tools~\cite{just2014mutants}. The study yielded a dataset of 
$676$ patches for $16$ different security vulnerability types, dubbed as Secbench. 
The vulnerability types are based on the OWASP Top $10$ of 
$2013$~\cite{oswap:2013} and OWASP Top $10$ of 
$2017$~\cite{oswap:2017}. Each test case of the dataset is a 
triplet: the commit before the patching (\emph{sha-p}), the commit responsible
for the patching (\emph{sha}), and the snippets of code that differ from one 
version to another (typically, called \emph{diffs})---where one 
can easily review the code used to fix the vulnerability. 

\textbf{Ponta et al. ($2019$)} tracked the \url{pivotal.io} website for 
vulnerabilities from $2014$ to $2019$. For each new vulnerability, 
the authors manually searched for references to commits involved in 
the patch on the National Vulnerability Database (NVD) website. 
However, $70\%$ of the vulnerabilities did not have any references 
to commits. Thus, the authors used their expertise to locate the 
commits in the repositories. This technique yielded a dataset of 
$624$ patches~\cite{10.1109/MSR.2019.00064} and $1282$ commits---one 
patch can have multiple commits assigned. To fit the dataset in our methodology, we located the 
first and last commits used to patch the vulnerability. For these 
cases, we used the GitHub API to retrieve the dates of 
the commits automatically. Then, for each patch, the group of commits was ordered 
from the oldest commit to the newest one. We assumed the last commit 
(newest one) as the fix (\emph{sha}) and the parent of the 
first commit (oldest commit) as the vulnerable version 
(\emph{sha-p}). 

In this study, we focus on computing the maintainability of the 
commits before and after the security patching to evaluate if its 
impact was positive, negative, or none. The $1300$ patches in the 
dataset were analyzed using the BCH toolset to calculate their 
maintainability reports. Due to the limitations of BCH (in particular, 
lack of language support and project size) and the presence of
floss-refactorings, $331$ patches were tossed---explained in 
more detail in Section~\ref{sec:main_analysis}.
The final dataset used in this 
paper comprises $969$ security patches from $260$ projects. We used the 
Common Weakness Enumeration (CWE) taxonomy to classify each vulnerability. 
For instance,  the \texttt{Fix CVE-2014-1608: mc\_issue\_attachment\_get SQL injection} 
(\texttt{00b4c17}\footnote{CVE-$2014$-$1608$ details available at 
\url{https://github.com/mantisbt/mantisbt/commit/00b4c17088fa56594d85fe46b6c6057bb3421102} 
(Accessed on \today)}) is a \emph{CWE-89: Improper Neutralization of Special Elements used 
in an SQL Command ('SQL Injection')}\footnote{CWE-$89$ details available at \url{https://cwe.mitre.org/data/definitions/89.html} 
(Accessed on \today)} according to the CWE taxonomy.
We were able to classify a total of $866$ patches using the CWE taxonomy:
the CWE's for $536$ patches were automatically scraped from the National 
Vulnerability Dataset (NVD); while the other $370$ patches were manually classified by 
the authors following the \emph{Research Concepts} CWE's list\footnote{Research Concepts list available at \url{https://cwe.mitre.org/data/definitions/1000.html}}. A total 
of $103$ patches were not classified because we were not able to map
the issue to any CWE with confidence due to the lack of quality
information on the vulnerability/patch.

\subsection{Security Patches vs. Regular Changes}
Previous studies attempted to measure the impact of regular changes 
on open-source software maintainability~\cite{HEGEDUS2018313}. 
However, there is no previous work focused on comparing the impact 
of security patches with regular changes on maintainability, only
with bug-fixes~\cite{10.1145/3133956.3134072}.
We analyze the maintainability of regular changes---changes not 
related to security patches---and, use them as a baseline.
The baseline dataset is generated from the security commits dataset, i.e.,
for each security commit in the dataset, we collect a random regular
change from the same project. We created two different baselines: 
\textit{random-baseline},
considering random changes and all their characteristics; and,
\textit{size-baseline}, considering also random changes
but with an approximate size as security patches---we argue 
that comparing changes with considerably different sizes may be unfair.

\subsubsection{Random-Baseline} 

As for the security patches, for each regular change, we  
need the commit performing the regular change (\emph{sha-reg}) and 
version of the software before the change (\emph{sha-reg-p}). A 
random commit from the same project is selected for each security patch, \emph{sha-reg}. The parent commit
of \emph{sha-reg} is the \emph{sha-reg-p}.

\subsubsection{Size-Baseline} 

For the size-baseline, we also need to obtain the regular change (\emph{sha-reg})
and the version of the software before the change (\emph{sha-reg-p}). 
First, our tool calculates the \emph{diff}
between the security patch and its parent.
Second, a random commit/regular change from 
the same project is selected, \emph{sha-reg}. The 
\emph{diff} between \emph{sha-reg} and its parent (\emph{sha-reg-p})
is calculated. Then, the regular change \emph{diff} is compared 
to the security patch \emph{diff}.
Due to the complexity of some patches, it was not possible 
to find patches with the exact same number of added and deleted lines. 
Thus, we looked for an approximation.

The pair of the regular change (\emph{sha-reg}) and its parent (\emph{sha-reg-p}) is accepted if the 
\emph{diff} size fits in the range size. This range widens every 
$10$ attempts to search for a change with an approximate size. We 
originate the regular changes from the security commits to ensure 
that differences in maintainability are not a consequence of 
characteristics of different projects.

\subsection{Bettter Code Hub}

SIG---the company behind BCH---has been helping business and technology 
leaders drive their organizational objectives by fundamentally improving 
the health and security of their software applications for more than $20$ years. 
The inner-workings of their SIG-MM model---the one behind BCH---are 
scientifically proven and certified~\cite{4335232,5609747,6113040,baggen2012}.

BCH checks GitHub codebases against $10$ maintainability 
guidelines~\cite{Visser:2016:OREILLY} that were empirically 
validated in previous work~\cite{Bijlsma:2012:FIR:2317098.2317124,8530041,8919169,8785997}.
SIG has devised these guidelines after many years of experience: analyzing more 
than $15$ million lines of code every week, SIG maintains the industry’s largest 
benchmark, containing more than $10$ billion lines of code across $200$+ technologies; SIG 
is the only lab in the world certified by TÜViT to issue ISO $25010$ certificates\footnote{Information available here: 
\url{https://www.softwareimprovementgroup.com/methodologies/iso-iec-25010-2011-standard/}}.
BCH's compliance criterion is derived from the requirements for 4-star 
level maintainability (cf. ISO $25010$)~\cite{5609747,6113040,baggen2012,Visser:2016:OREILLY}.
SIG performs the threshold calibration
yearly on a proprietary data set to satisfy the requirements of TUViT to
be a certified measurement model.

As BCH, other tools also perform code analysis for similar 
metrics. Two examples are Kiuwan and SonarCloud. However, Kiuwan does
not provide the full description of the metrics it measures; and, SonarCloud
although it provides a way of rating software maintainability, the variables 
description of their formula are not available. Both analyze less maintainability
guidelines than BCH and do not have their inner workings fully and publicly described.

\subsection{Maintainability Analysis}\label{sec:main_analysis}

In this research, we follow a very similar methodology to the one 
presented in previous work on the maintainability of energy-oriented 
fixes~\cite{8919169}. The inner workings of BCH were proposed
originally in 2007~\cite{4335232} and suffered refinements later~\cite{5609747,6113040,baggen2012}. 
As said before, the web-based source code 
analysis service \emph{Better Code Hub} (BCH) is used to collect the 
maintainability reports of the patches of each project. 
Table~\ref{tab:guidelines} presents the $10$ guidelines proposed
by BCH's authors for delivering software that is not difficult to
maintain~\cite{Visser:2016:OREILLY} and, maps each guideline to the 
metric calculated by BCH. These guidelines are calculated using the 
metrics presented in~\cite{criteria:2017} and are also briefly explained in Table~\ref{tab:guidelines}. 
During each guideline evaluation, the tool determines the compliance 
towards one guideline by establishing limits for the percentage of 
code allowed to be in each of the $4$ risk severity levels
(\emph{low risk}, \emph{medium risk}, \emph{high risk}, and 
\emph{very high risk}). If the project does not violate those 
thresholds, then the BCH considers that the code is compliant with 
a guideline. These thresholds are determined by BCH using their own
data/experience---using open-source and closed software systems. If 
a project is compliant with a guideline, it means that it is at 
least $65\%$ better than the software used by BCH to calculate the 
thresholds\footnote{Check the answer to \emph{How can I adjust the 
threshold for passing/not passing a guideline?} at
\url{https://bettercodehub.com/docs/faq} (Accessed on \today{})}.

\begin{table}[h]
\centering
\scriptsize
	\caption{Guidelines to produce maintainable code.}
\begin{tabular}{L{2cm}L{4.5cm}L{4cm}}

\toprule
\textbf{10 Guidelines} & \textbf{Description} & \textbf{Metric}\\
\midrule

\textbf{Write Short Units of Code} & Limit code units to $15$ LOCs because smaller
 units are easier to understand, reuse and test them & \textbf{Unit Size:} \% of 
 LOCs within each unit~\cite{criteria:2017} \\\midrule

\textbf{Write Simple Units of Code} & Limit branch points to $4$ per unit because
it makes units easier to test and modify & \textbf{McCabe Complexity:} \# of decision 
points~\cite{1702388,criteria:2017}\\\midrule

\textbf{Write Code Once} & Do not copy code because bugs tend to replicate at
multiple places (inefficient and error-prone) & \textbf{Duplication:} \% of redundant 
LOCs~\cite{criteria:2017}\\\midrule

\textbf{Keep Unit Interfaces Small} & Limit the number of parameters to at most
$4$ because it makes units easier to understand and reuse & \textbf{Unit Interfacing:} 
\# of parameters defined in a signature of a unit~\cite{criteria:2017} \\\midrule

\textbf{Separate Concerns in Modules} & Avoid large modules because changes in
loosely coupled databases are easier to oversee and execute & \textbf{Module Coupling:} \# of
incoming dependencies~\cite{criteria:2017} \\\midrule

\textbf{Couple Architecture Components Loosely} & Minimize the amount of code
within modules that are exposed to modules in other components & \textbf{Component Independence:} 
\% of code in modules classified as hidden~\cite{criteria:2017}\\\midrule

\textbf{Keep Architecture Components Balanced} & Balancing the number of
components ease locating code and allow for isolated maintenance & \textbf{Component Balance:} 
Gini coefficient to measure the inequality of distribution between components~\cite{criteria:2017} \\\midrule

\textbf{Keep your code base Small} & Reduce and avoid the system size because
small products are easier to manage and maintain & \textbf{Volume:} \# of LOCs converted 
to man-month/man-year~\cite{criteria:2017} \\\midrule

\textbf{Automate Tests} & Test your code base because it makes development
predictable and less risky & \textbf{Testability:} Ratings aggregation $-$ unit 
complexity, component independence and volume~\cite{Visser:2016:OREILLY}
 \\\midrule

\textbf{Write Clean Code} & Avoid producing software with code smells because
it is more likely to be maintainable in the future & \textbf{Code Smells:} 
\# of Occurrences~\cite{Visser:2016:OREILLY} (e.g., magic constants and long 
identifier names) \\
\bottomrule
\end{tabular}
\label{tab:guidelines}
\end{table}

\begin{figure}[h]
  \centering  
  \includegraphics[width=0.8\textwidth]{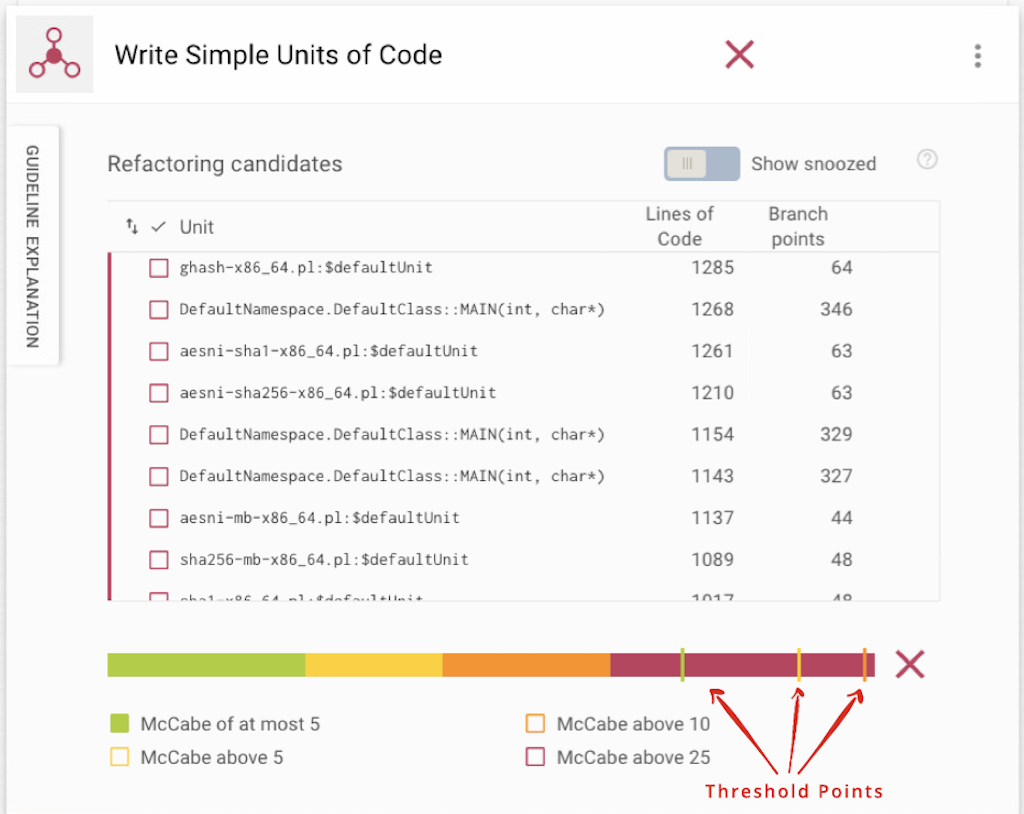}
  \caption{Maintainability report of OpenSSL's CVE-$2016$-$6304$ 
  vulnerability patch for the guideline \emph{Write Simple Units 
  of Code} provided by \emph{Better Code Hub}. This version of 
  OpenSSL does not comply with the guideline in the example since 
  the bars do not reach the threshold points. This example only 
  complies with $\sfrac{1}{10}$ guidelines (\emph{Write Clean 
  Code}).}
  \label{fig:bchrep}
\end{figure}

Figure~\ref{fig:bchrep} shows an example of the report provided by BCH 
for a project after finishing its evaluation. The example
refers to the OpenSSL CVE-$2016$-$6304$ vulnerability patch---
as described by Section~\ref{sec:motivation}. This version of 
OpenSSL only complies with $1$ out of $10$ guidelines: \emph{Write 
Clean Code}.

SIG defines \emph{Units} as the smallest groups of code that can be 
maintained and executed independently~\cite{Visser:2016:OREILLY} 
(e.g., methods and constructors in Java). One of the guidelines with 
which the project does not comply is the one presented in the report 
(cf. Figure~\ref{fig:bchrep}): \emph{Write Simple Units of Code}. BCH 
analyzes this guideline based on the McCabe 
Complexity~\cite{1702388} to calculate the number of branch points 
of a method. The bar at the bottom of the figure represents the top 
$30$ units that violate the guideline, sorted by severity. The 
different severities of violating the guideline are indicated using 
colors, and there is a legend to help interpret them. The green 
bar represents the number of compliant branch points per unit 
(\emph{at most $5$}), i.e., the number of units are compliant with 
ISO $25010$~\cite{iso:2011}. Yellow, orange, and red bars represent 
units that do not comply with medium (\emph{above $5$}), high 
(\emph{above $10$}) and very high (\emph{above $25$}) severity 
levels. In the bar, there are marks that pinpoint the compliance 
thresholds for each severity level. If the green mark is somewhere 
in the green bar, it is compliant with a low severity level.

\begin{figure}[h]
 	\centering 	
	\includegraphics[width=0.5\linewidth]{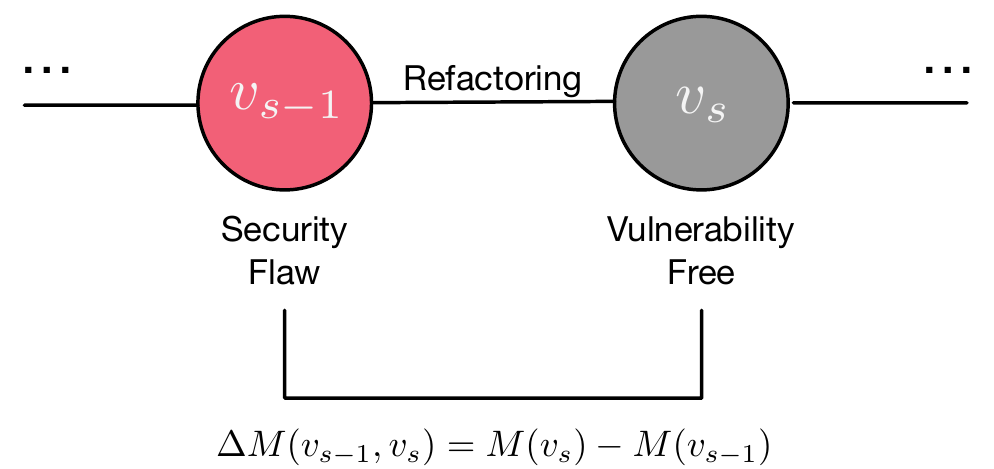}
 	\caption{Maintainability difference for security commits.}
	\label{fig:commit}
\end{figure}

Aiming to analyze the impact of security patches, we use BCH to compute 
the maintainability of two different versions of the project 
(cf. Figure~\ref{fig:commit}):

\begin{itemize}
	\item $v_{s-1}$, the version containing the security flaw, i.e., 
	before the patch (\emph{sha-p});
	\item $v_{s}$, the version free of the security flaw, i.e., 
	after the patch (\emph{sha});
\end{itemize}

Security patches can be performed through one commit (single-commit); 
several consecutive commits (multi-commits); or, commit(s) interleaved with more 
programming activities (floss-refactoring). Only $10.7\%$ of the data points 
of our dataset involve more than one commit, the other
$89.3\%$ of the cases are single-commit patches. To mitigate the impact of 
floss-refactorings, we extracted and manually inspected a random sample with 
$25\%$ of security patches from each dataset. From this sample, we identified 
$23$ floss-refactorings. Most floss-refactoring patches include many changes 
making it difficult to understand which parts involve the security patch. 
Although we suspect that more floss-refactorings may occur, we argue that 
they occur in a small portion of the data.

Due to BCH limitations, in particular, lack of language support 
and project size by BCH, $308$ data points were not analyzed and automatically 
disregarded from our study. After performing the BCH analysis and the 
maintainability calculations, we found the following 
limitations regarding two of BCH's guidelines:

1) For projects with large codebases, the results calculated for 
the \emph{Keep Your Codebase Small} guideline were way above the 
limit set by BCH ($20$ person-years). We suspect this threshold 
may not be well-calibrated, and hence biasing our results. Thus, 
we decided not to consider this guideline in our research. 

2) The 
\emph{Automated Tests} guideline was also not considered since the 
tool does not include two of the most important techniques to 
security testing: vulnerability scanning and penetration testing. 
Instead, it only integrates unit testing.

The BCH tool does not compute the final score that our study needs 
to compare maintainability amongst different project versions. We 
follow previous work on measuring the impact of 
energy-oriented patches~\cite{8919169}. Cruz et al. ($2019$) 
proposed an equation to capture the distance between the current 
state of the project and the standard thresholds calculated by the 
BCH based on the insights provided in~\cite{Olivari:2018}. The 
equation provided in~\cite{8919169} considers that the size of 
project changes do not affect the maintainability, and that the 
distance to lower severity levels is less penalized than to the
thresholds in high severity levels.


Given the violations for the BCH guidelines, the maintainability 
score is computed $M(v)$ as follows:

\begin{equation}
    M(v) = \sum_{g \in G}^{} M_{g}(v)
\end{equation}

\noindent
where $G$ is the group of maintainability guidelines from BCH
(Table~\ref{tab:guidelines}) and $v$ is the version of the software 
under evaluation. $M(v) < 0$ indicates that version $v$ is violating 
(some of) the guidelines, while $M(v) > 0$ indicates that version 
$v$ is following the BCH guidelines. The maintenance for the 
guideline $g$, $M_g$ for a given version of a project is computed as 
the summation of the compliance with the maintainability guideline 
for the given severity level (medium, high, and very high).
The compliance for a severity level is calculated based on previous 
work, which calculates the number of lines of code that comply and 
not comply with the guideline at a given severity 
level~\cite{8919169}. In our analysis, we compute the difference of 
maintainability between the security commit ($v_{s}$) and its parent 
commit ($v_{s-1}$), as illustrated in Figure~\ref{fig:commit}. Thus, 
we can determine which patches had a positive, negative, or null 
impact on the project maintainability. 

%
%
%
%

\subsection{Statistical Validation}\label{sec:statsval}
To validate the maintainability differences in different groups of 
commits (e.g., baseline and security commits), we use the Paired 
Wilcoxon signed-rank test with the significance level $\alpha = 
0.05$~\cite{10.2307/3001968}. In other words, we test the null 
hypothesis that the maintainability difference between pairs of 
versions $v_{s-1}$, $v_s$ (i.e., before and after a security commit) 
come from the same distribution. Nevertheless, this test has a 
limitation: it does not consider the groups of commits with 
zero-difference maintainability. In $1959$, Pratt improved the 
test to solve this issue, making the test more robust. Thus, 
we use a version of the Wilcoxon test that 
incorporates the cases where maintainability is equal to 
zero~\cite{10.2307/2282543}. The Wilcoxon test requires a 
distribution size of at least $20$ instances. To understand the 
effect-size, as advocated by the Common-language effect 
sizes, we compute the mean difference, the median of 
the difference, and the percentage of cases that reduce 
maintainability~\cite{graw:1992}.

\section{Results \& Discussion}\label{sec:results}

This study evaluates a total of $969$ security patches and 
$969$ regular changes 
from $260$ distinct open-source projects. 
This section 
reports and discusses the results for each research question.

\textit{\textbf{RQ1: What is the impact of security patches on the
maintainability of open-source software?}} In \emph{RQ1}, we 
report and discuss the impact of patches on open-source software 
maintainability under four groups: guideline, overall score, 
severity and programming language.

\begin{figure}[htp]
     \begin{adjustwidth}{-1cm}{-1cm}  
  	\centering
  	\includegraphics[width=0.9\textwidth]{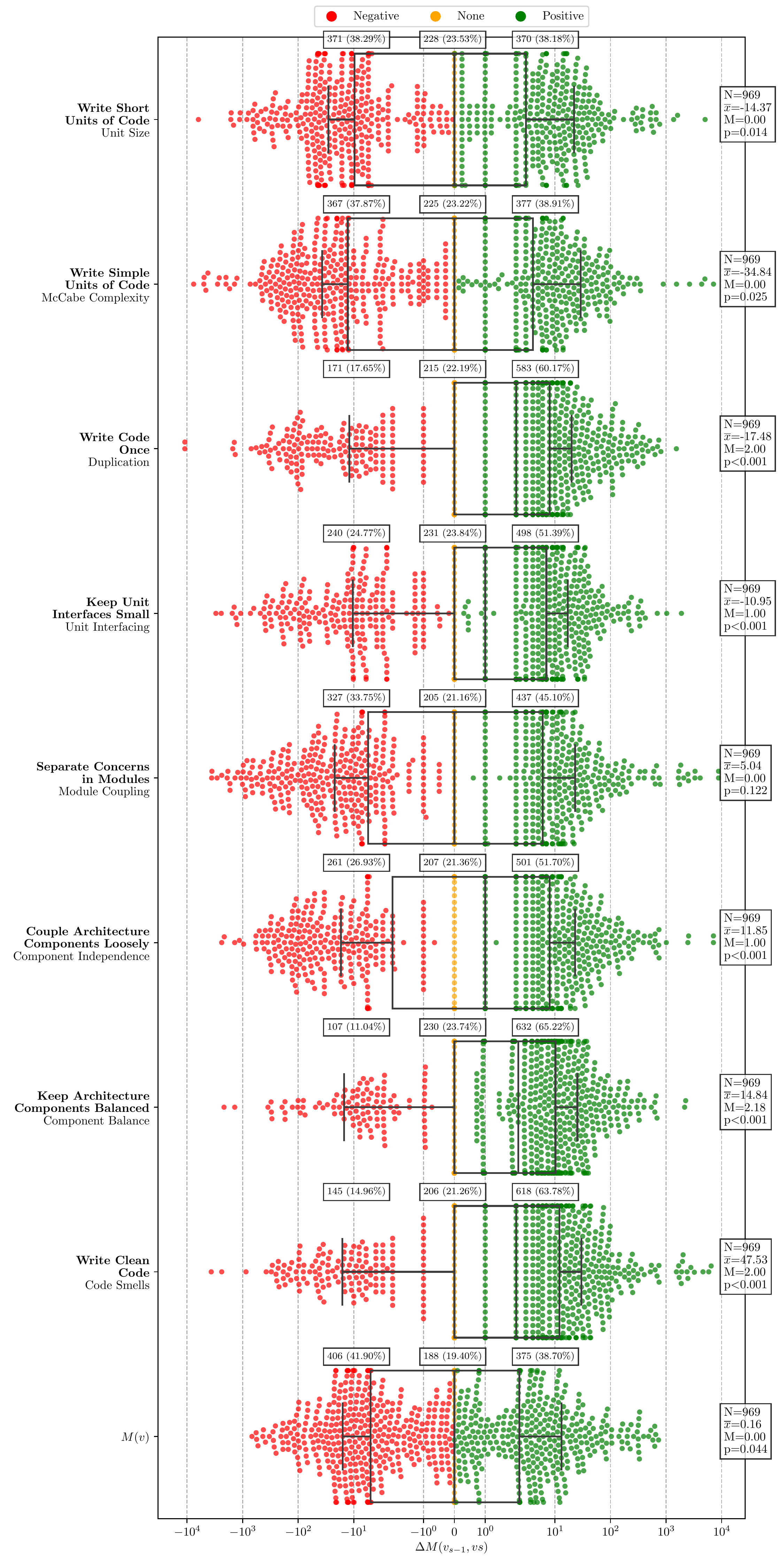}
    \end{adjustwidth}
	
  	\caption{Impact of the security patches per guideline and overall mean, $M(v)$. 
    Each point of the plot represents the impact of a security patch on software 
    maintainability. Red means the impact was negative, i.e., the patch 
    harmed maintainability. Yellow means the patch did not have any 
    kind of impact on maintainability. Green means the impact was 
    positive, i.e., the patch improve software maintainability. For instance, in the 
    \emph{Write Short Units of Code} guideline, $38.29\%$ of 
    security patches harmed software maintainability; $23.53\%$ of 
    security patches had no impact on maintainability; and, 
    $38.18\%$ of security patches improved software maintainability.}
 	\label{fig:guidelines}	
 \end{figure}

\textbf{Guideline/Metric:} Each patch performs a set of changes
on the software's source code. These changes may have a 
different impact on the guidelines/metrics used to measure software 
maintainability. Figure~\ref{fig:guidelines} shows the impact of 
security patches on each guideline individually and the average 
impact on all guidelines together ($M(v)$). Under each guideline, it 
is stated the metric used for the calculations. For instance, for the 
\emph{Write Short Units of Code} guideline, the metric used is 
\emph{Unit Size}. Table~\ref{tab:guidelines} describes in more 
detail the metrics behind the guidelines. For each type of 
guideline, a swarm plot is presented to show the variability/dispersion 
of the results alongside the number of absolute and 
relative cases of each impact. Next to each type of guideline, it is 
presented the mean ($\overline{x}$) and median (M) of the 
maintainability difference and the p-value resulting from the Paired 
Wilcoxon signed-rank test. $M(v)$ is not a guideline but rather the 
average impact of all guidelines. Each point of the plot represents 
the impact of a security patch on software maintainability. Red 
means the impact was negative, i.e., the patch harmed 
maintainability. Yellow means the patch did not have any kind of 
impact on maintainability. Green means the impact was positive, i.e., 
the patch improved software maintainability.

 \begin{figure}[htp]
  	\centering 	 	\includegraphics[width=0.65\textwidth]{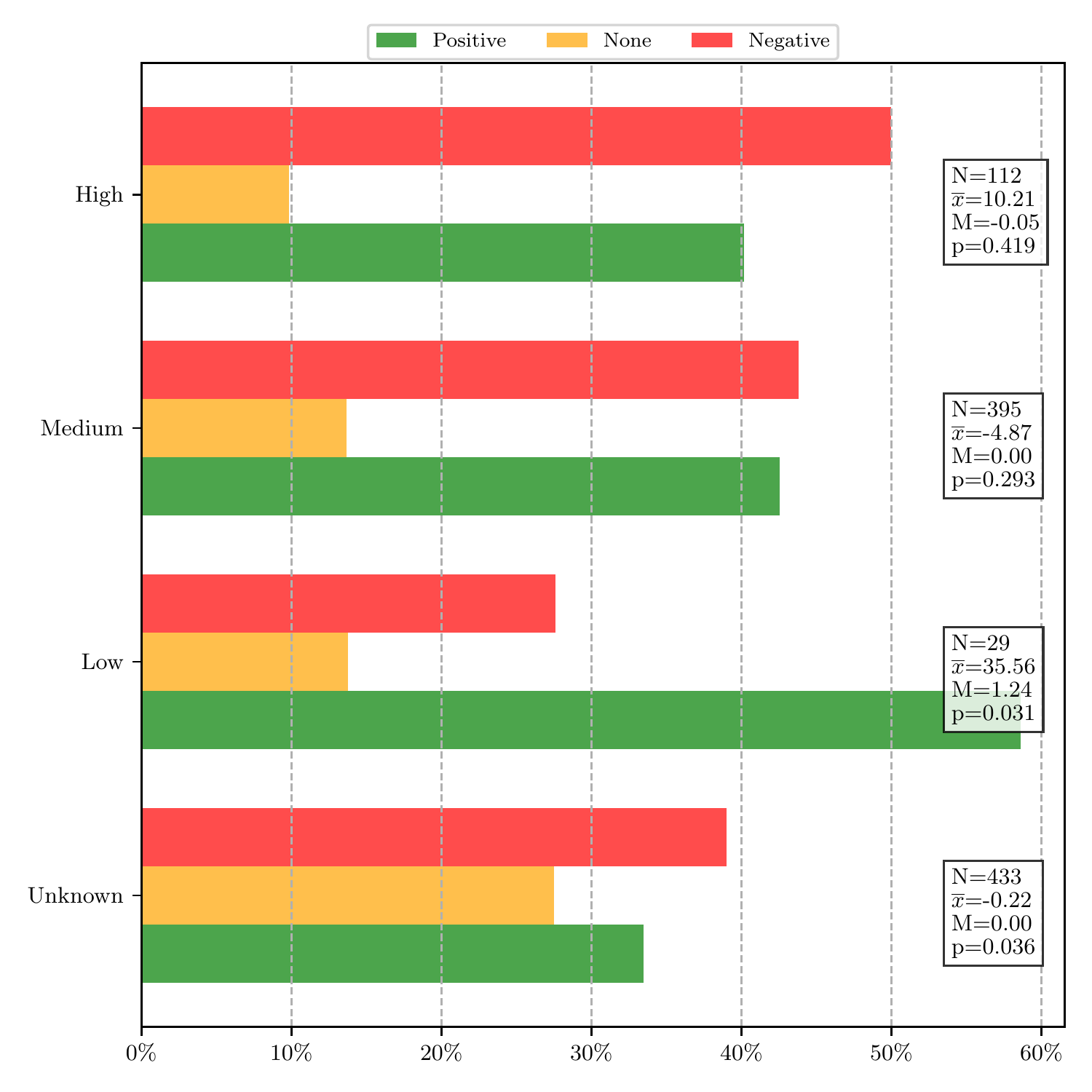}
  	\caption{Maintainability difference by vulnerability severity.}
 	\label{fig:severity}
 \end{figure}

 Regarding the impact of security patches per guideline, we 
 observe that $38.7\%$ of the security patches have positive impact
 on software maintainability. However, we also 
 see that patching vulnerabilities have a very significant number of 
 negative cases per guideline---between $10\%$ and $40\%$. 
 \emph{Write Short Units of Code} ($38.3\%$), \emph{Write Simple 
 Units of Code} ($37.9\%$), and \emph{Separate Concerns in Modules} 
 ($33.8\%$) seem to be the most negatively affected guidelines. This 
 may imply that developers, when patching vulnerabilities, have a hard 
 time designing/implementing solutions that continue to respect the 
 limit bounds of branch points and function/module sizes that are 
 recommended by coding practices. Still, on respecting bound limits, 
 developers also seem not to consider the limit of $4$ parameters per 
 function for the \emph{Keep Unit Interfaces Small} guideline 
 required by BCH, in $24.8\%$ of the cases. This guideline is usually
 violated when the patch requires to input new information to a 
 function/class, and developers struggle to use the \emph{Introduce 
 Parameter Object} patch pattern. Results do not provide statistical 
 significance to the \emph{Separate Concerns in Modules} guideline, 
 i.e., results should be read carefully.

Software architecture is also affected while patching 
vulnerabilities. Both \emph{Couple Architecture Component Loosely} 
and \emph{Keep Architecture Components Balanced} guidelines suffer a 
negative impact of $26.9\%$ and $11.0\%$, respectively. Component 
independence and balance are important to make it easier to find the 
source code that developers want to patch/improve and to understand 
how the high-level components interact with others. However, results 
may imply that developers forget to use techniques such as 
encapsulation to hide implementation details and make the system 
more modular.

The \emph{Write Code Once} guideline results show that duplicated 
code increased in $17.7\%$ ($171$/$969$) of the patches. Software systems 
typically have $9\%$-$17\%$ of cloned code~\cite{5773403}. Previous work showed a correlation between code smells and code
duplication~\cite{7476787} which may also be reflected in the 
\emph{Write Clean Code} guideline results. BCH reported new code 
smells for $15.0\%$ ($145$/$969$) of the patches, which according to previous work, 
may be the source of new software vulnerabilities~\cite{8819456,10.1145/3133956.3134072}
capable of harming the market value and economy of companies~\cite{4267025}.
Developers should never reuse code by copying and pasting 
existing code fragments. Instead, they should create a method and call 
it every time needed. The \emph{Extract Method} refactoring 
technique solves many duplication problems. This makes spotting and solving 
the issue faster because you only need to fix the method used
instead of locating and fixing the issue multiple times.
Clone detection tools can also help in locating the issues.

\textbf{Overall Score ($M(v)$):} 
Although overall patching vulnerabilities has a less negative impact
on software maintainability guidelines, this is not reflected in the 
average impact of all guidelines ($M(v)$) as we can see in 
Figure~\ref{fig:guidelines}. Remember that each point of the plot 
represents the impact of a security patch on software 
maintainability. Red means the impact was negative, i.e., the patch 
harmed maintainability. Yellow means the patch did not have any kind of 
impact on maintainability. Green means the impact was positive, i.e., the patch improved software maintainability.
The $M(v)$ plot shows that $406$ ($41.9\%$) cases have a negative impact 
on software maintainability. While $188$ ($19.4\%$) cases 
have no impact at all, and $375$ ($38.7\%$) have a positive impact on 
software maintainability. The larger number of negative cases may be 
explained by guidelines with higher concentrations of negative 
cases with higher amplitudes, such as \emph{Write Short Units of 
Code}, \emph{Write Simple Units of Code} and \emph{Separate Concerns 
in Modules}---more red points on the left, being $0$ the reference point.
The resulting p-value of the Paired Wilcoxon signed-rank test for $M(v)$ 
is $0.044$ (cf. Figure~\ref{fig:guidelines}). Since the p-value is 
below the significance level of $0.05$, we argue that security patches 
may have a negative impact on the maintainability of open-source software.

\textbf{Severity:} Some of the vulnerabilities are identified with 
\emph{Common Vulnerabilities and Exposure} (CVE) entries. We 
leveraged the \emph{National Vulnerability Database} (NVD) website 
to collect their severity levels. In total, we retrieved severity 
scores for $536$ vulnerabilities: $112$ \emph{High}, $395$ 
\emph{Medium} and $29$ \emph{Low}. Figure~\ref{fig:severity} 
presents the impact of security patches per severity level on the 
maintainability of open-source software. We observe that patches for 
\emph{High} ($50.0\%$) and \emph{Medium} ($43.8\%$) severity 
vulnerabilities hinder more the maintainability of software than 
\emph{Low} ($27.6\%$) severity vulnerabilities. Again, patches have 
a considerable negative impact on software maintainability---between 
$20\%$ and $50\%$. Statistical significance was retrieved only for 
\emph{Low} severity vulnerabilities, i.e., \emph{Low} severity 
vulnerabilities may have more cases where software maintainability was improved than the 
other severity levels. However, results should not be disregarded 
because they somehow confirm the 
assumption that higher severity vulnerabilities patches may have a 
more negative impact on maintainability, i.e., high/medium 
severity vulnerabilities may need more attention than low severity while 
patching.

\begin{figure}[htp]
  \centering
  \includegraphics[width=0.6\textwidth]{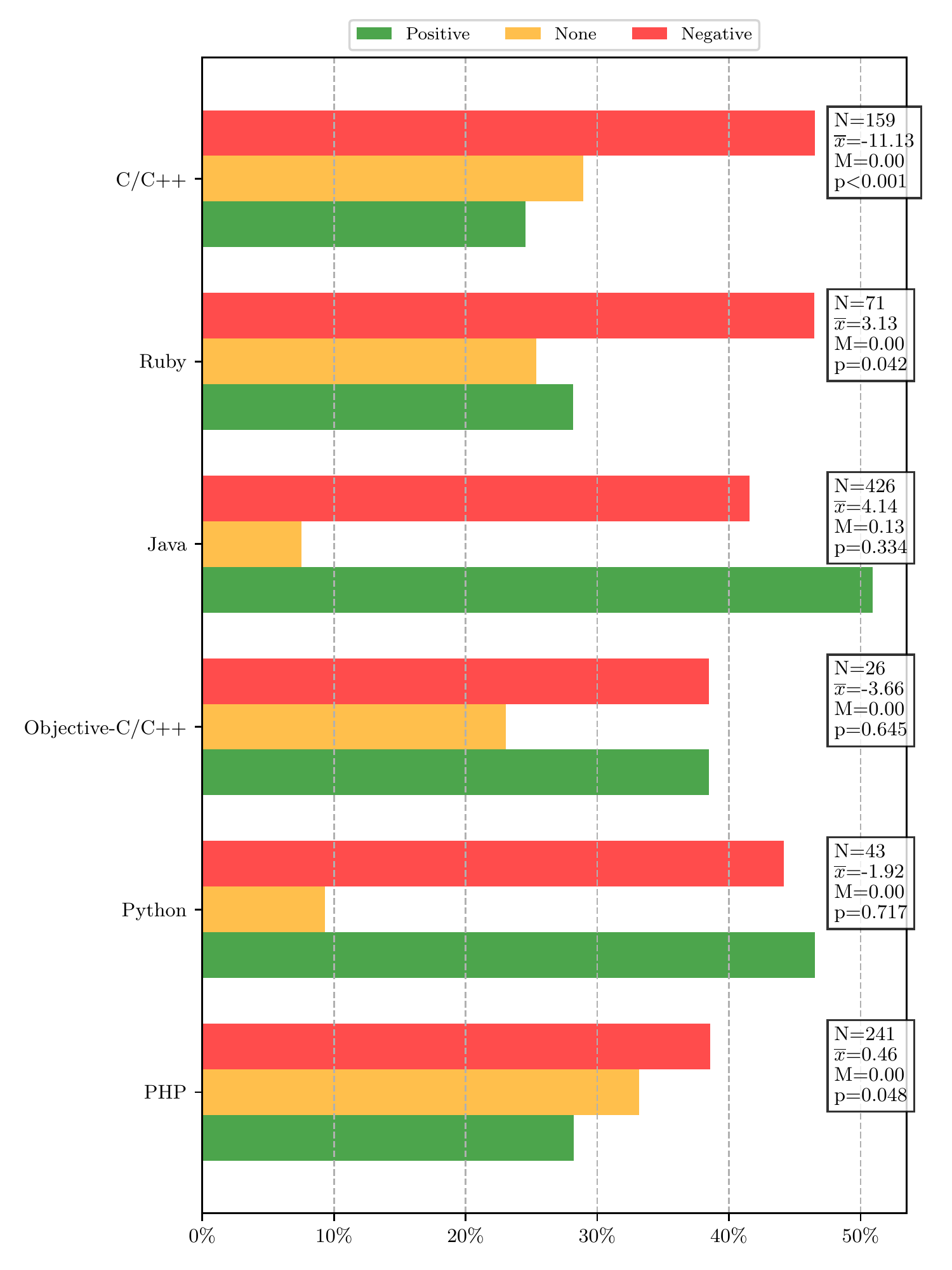}
  \caption{Maintainability difference by programming language.}
  \label{fig:lang_main}    
\end{figure}

\begin{figure*}[htp]
  \centering
  \subfigure[Maintainability difference by first-level weaknesses from the 
  \textit{Research Concepts} list on Common Weakness Enumeration (CWE)]{
  \includegraphics[scale=0.4]{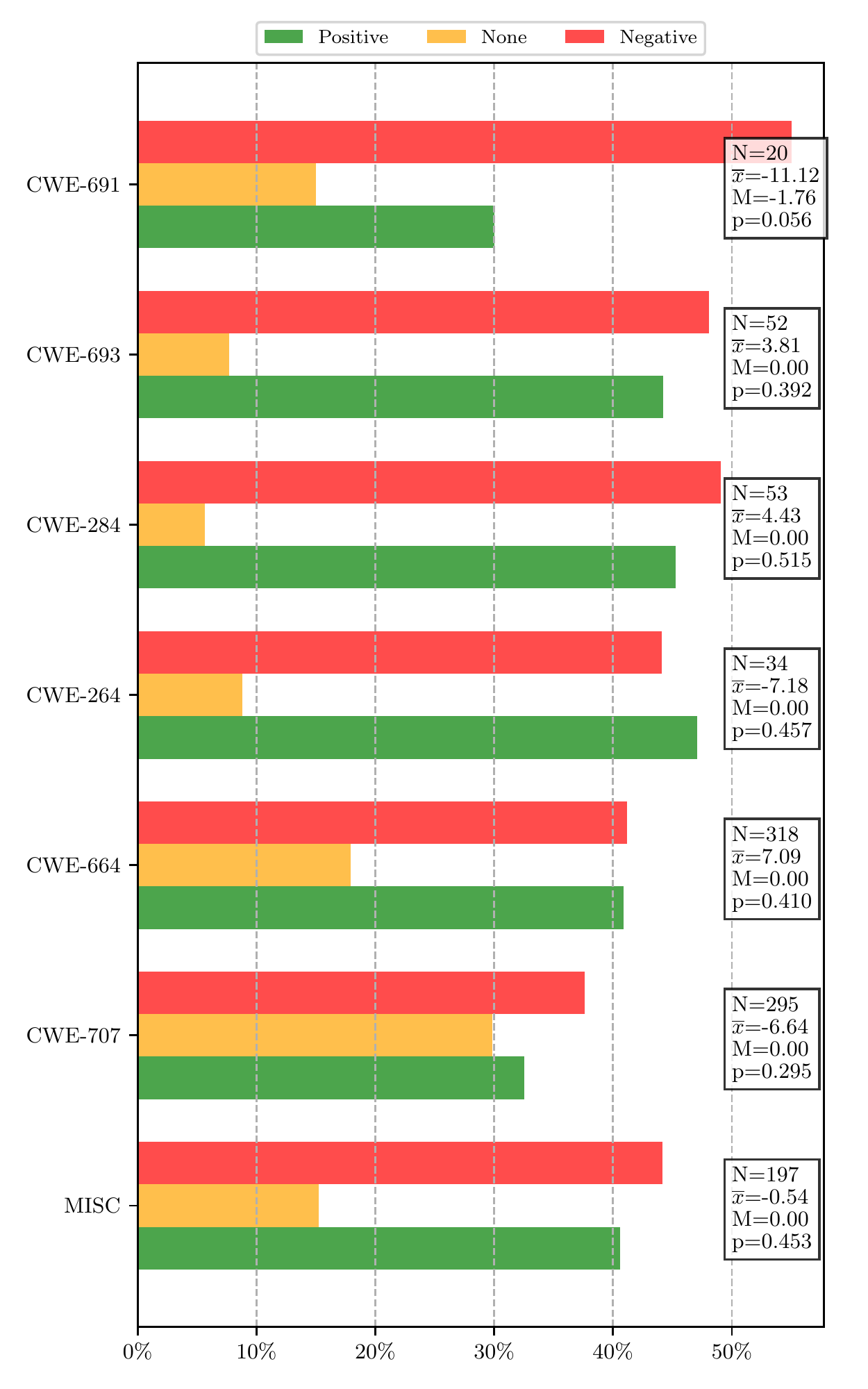}}\quad\quad
  \subfigure[Maintainability difference by sub-weaknesses of the 
  \textit{Improper Neutralization} Weakness (CWE-707)]{
  \includegraphics[scale=0.4]{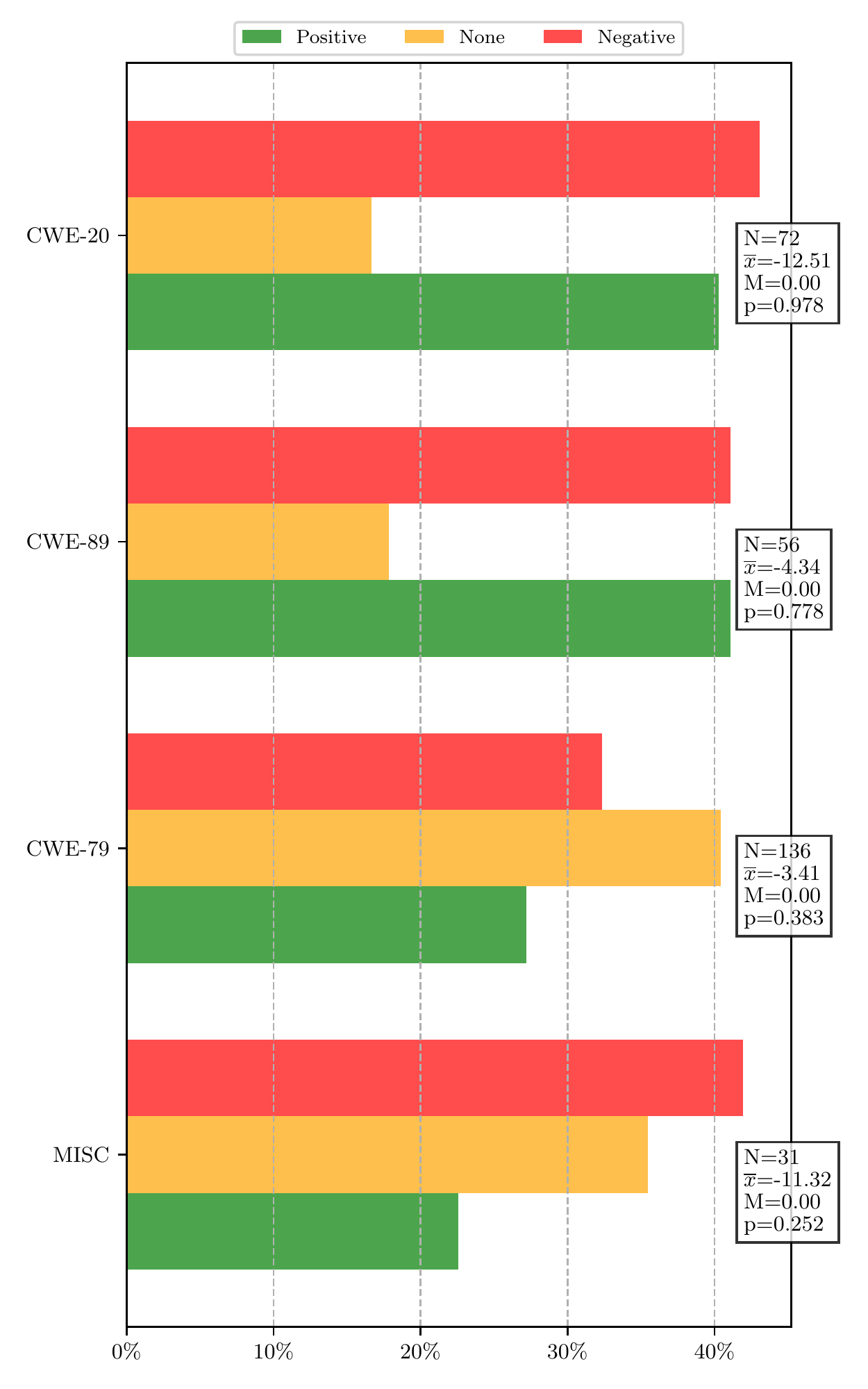}}\quad\quad
  \subfigure[Maintainability difference by sub-weaknesses of the 
	\textit{Improper Control of a Resource Through its Lifetime} Weakness (CWE-664)]{
	\includegraphics[scale=0.4]{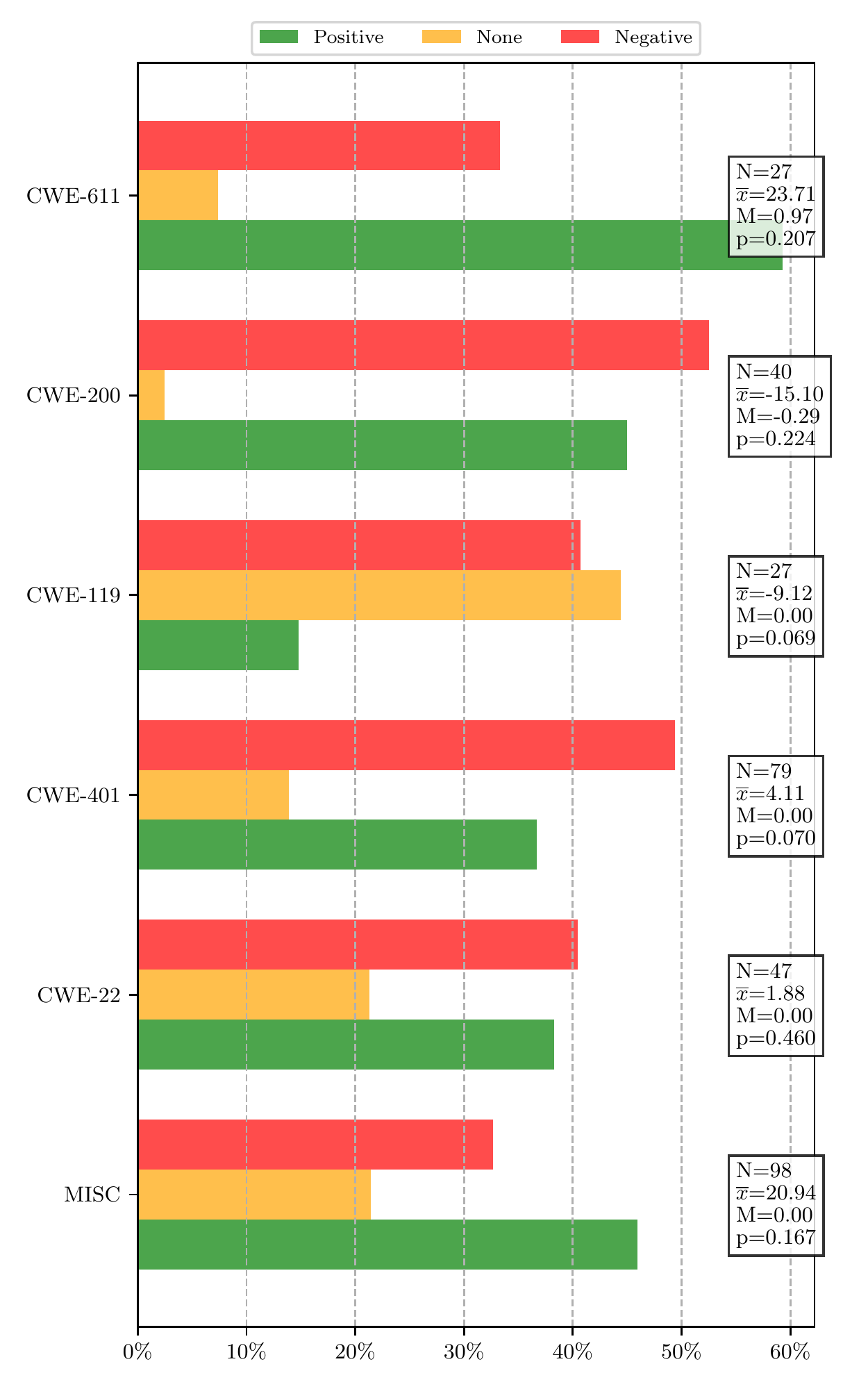}}
    \caption{Maintainability difference per weakness.}
	\label{fig:pat}
\end{figure*}

\textbf{Programming Language:} The impact on software maintainability per programming 
language was also analyzed (Figure~\ref{fig:lang_main}). We restrict 
this analysis to programming languages with at least $20$ data points, as this 
is a requirement for the hypothesis tests. Thus, we compare the results for 
\texttt{C/C++}, \texttt{Ruby}, \texttt{Java}, \texttt{Objective C/C++}, 
\texttt{Python} and \texttt{PHP}, leaving \texttt{Groovy} out of the analysis.
\texttt{C/C++}, \texttt{Ruby} and 
\texttt{PHP} are the programming languages with worse 
impact on maintainability, i.e., with the highest number of negative cases ($46.5\%$, $46.5\%$ 
and $38.6\%$, respectively). \texttt{Java} and \texttt{Python} 
seem to be less affected by patching, i.e., integrating a larger amount of cases with positive 
impact on maintainability ($50.9\%$ and $46.5\%$, respectively). But overall 
languages have a considerable amount of cases that negatively impact maintainability---between $35\%$ to $50\%$---which 
confirms the need for better/more secure programming languages. 
Statistical significance was only retrieved for the \texttt{C/C++} 
($p = 2.24$x$10^{-05}$), \texttt{Ruby} ($p = 0.041$) and \texttt{PHP} 
($p = 0.048$) languages. Yet, data reports very interesting hints on the impact 
of programming languages on security patches.

We expected the negative impact for programming languages on
maintainability to be more severe, as arguably, poor design of programming
languages for security and the lack of best practices application by developers lead to more buggy/vulnerable
code~\cite{Ray:2017:LSP:3144574.3126905,2019arXiv190110220B}. However,
Figure~\ref{fig:lang_main} shows that only \emph{C/C++} and \emph{Ruby} have 
a significant negative impact approximate to $50\%$ on 
maintainability. We suspect that these values are
the result of project contributions policies (e.g., coding standards). In our dataset, 
$9$/$10$ projects with more contributors follow strict contribution 
policies for code standards.

\textbf{Summary:} Results show that developers may have a
hard time following the guidelines and, consequently, hinder
software maintainability while patching vulnerabilities; and that different levels of
attention should be paid to each guideline. For instance,
\emph{Write Simple Units of Code} and \emph{Write Short Units of Code}
guidelines are the most affected ones. No statistical significance was observed
for \emph{Separate Concerns in Modules}.
As shown in Figure~\ref{fig:guidelines}, there is statistical significance ($p=0.044 < 0.05$) 
to support our findings: \textbf{security patches
may have a negative impact on the maintainability of open-source software}. Therefore, tools 
such as BCH should be integrated into the CI/CD pipelines
to help developers evaluate the risk of patches of hindering software maintainability---alongside
Pull Requests/Code Reviews. Different severity
vulnerabilities may need different levels of attention---high/medium 
vulnerabilities need more attention (cf. Figure~\ref{fig:severity}). However, 
statistical significance was only observed for low severity vulnerabilities. Better 
and more secure programming languages are needed. We observed statistical 
significance for \emph{C/C++}, \emph{Ruby} and \emph{PHP} that support that 
security patches in those languages may hinder software maintainability (cf. Figure~\ref{fig:lang_main}).

\textit{\textbf{RQ2: Which weaknesses are more likely to
affect open-source software maintainability?}}
In \emph{RQ2}, we report/discuss the impact of security patches on
software maintainability per weakness (CWE). We use the weakness definition
and taxonomy proposed by the \emph{Common Weakness Enumeration} (cf. Section~\ref{sec:motivation}).
Figure~\ref{fig:pat} shows three different charts. Figure \emph{\ref{fig:pat}-a}, presents
the impact of the $969$ patches grouped by the first level weaknesses from
the \emph{Research Concepts}\footnote{Research Concepts 
 is a tree-view provided by the Common Weakness Enumeration (CWE) website 
 that intends to facilitate research into weaknesses. It is organized 
 according to 
 abstractions of behaviors instead of how they can be detected, 
 their usual location in code, and when they are introduced in the 
 development life cycle. The list is available here: \url{https://cwe.mitre.org/data/definitions/1000.html}
} list. While the Figures \emph{\ref{fig:pat}-b} and
\emph{\ref{fig:pat}-c} present the impact on maintainability for lower levels of 
weaknesses for the most prevalent weaknesses in Figure \emph{\ref{fig:pat}-a}:
\emph{Improper Neutralization} (CWE-707) 
and \emph{Improper Control of a Resource 
Through its Lifetime} (CWE-664), respectively.

In Figure~\ref{fig:pat}-\emph{a}, there is no clear evidence of the impact on 
maintainability per weakness. Yet, it is important to note that
overall there is a very considerable number of cases that hinder
maintainability---between $30\%$ and $60\%$. The CWE-707 and CWE-664 
weaknesses integrate the higher number of cases compared to the remaining 
ones: $295$ ($30.4\%$) data points and $318$ ($32.8\%$) data points, respectively. 
Thus, we present an analysis of their sub-weaknesses on 
Figure~\ref{fig:pat}-\emph{b} and Figure~\ref{fig:pat}-\emph{c}, respectively. 

Results shows that patching vulnerabilities may hinder 
the maintainability of open-source software in $4$ different sub-weaknesses: 
\emph{Improper Input Validation (CWE-20)}, \emph{Information Exposure 
(CWE-200)}, \emph{Missing Release of Memory after Effective 
Lifetime (CWE-401)} and \emph{Path Traversal (CWE-22)}. Results also show that 
software maintainability is less negatively impacted when patching 
\emph{Improper Restriction of XML External Entity Reference (CWE-611)}.  

The impact of a patch depends on its complexity, i.e., if the patch
adds complexity to the code base, it is probably affecting the software
maintainability. \emph{Cross-Site Scripting (CWE-79)} and 
\emph{Improper Restriction of Operations within the Bounds of 
a Memory Buffer (CWE-119)} patches endure more
cases with no impact on the open-source software maintainability. \emph{SQL Injection
(CWE-89)} patches equally hinder and improve software maintainability.
These patches usually follow the same complexity as the CWE-79 patches.
However, the three weaknesses have a considerable amount of cases that hinder
the software maintainability---$32.4\%$, $40.7\%$ and, $41.1\%$, respectively---which
should not be happening. Typically, CWE-79 vulnerabilities do not need extra lines 
to be fixed, as shown in Listing~\ref{lst:fix}---one simple 
\texttt{escape} function patches the issue. On the same type of fix,
CWE-199 vulnerabilities may also be fixed without adding new source code
(e.g., replacing the \texttt{strncpy} function with a more secure one 
\texttt{strlcpy} that checks if the buffer is null-terminated). However,
some buffer overflows may be harder to fix and lead to more complex 
solutions (e.g., 
CVE-$2016$-$0799$\footnote{CVE-$2016$-$0799$ patch details available at 
\url{https://github.com/openssl/openssl/commit/9cb177301fdab492e4cfef376b28339afe3ef663}
(Accessed on \today{})}). 
As CWE-199 weaknesses, \emph{Missing Release of Memory after Effective 
Lifetime (CWE-401)} can also be the cause of Denial-of-Service attacks and difficult to 
patch since it usually requires adding complexity to the program (cf. Section~\ref{sec:motivation}). 

\textbf{Summary:}
Although results did not yield statistical significance, we show preliminary evidence that 
researchers and developers ought to pay more attention to maintainability when fixing the 
following types of weaknesses: \emph{Improper Input Validation (CWE-20)}, \emph{Information Exposure
(CWE-200)}, \emph{Missing Release of Memory after Effective
Lifetime (CWE-401)} and \emph{Path Traversal (CWE-22)}.

\begin{figure}[htp]
  \centering
  \includegraphics[width=0.8\textwidth]{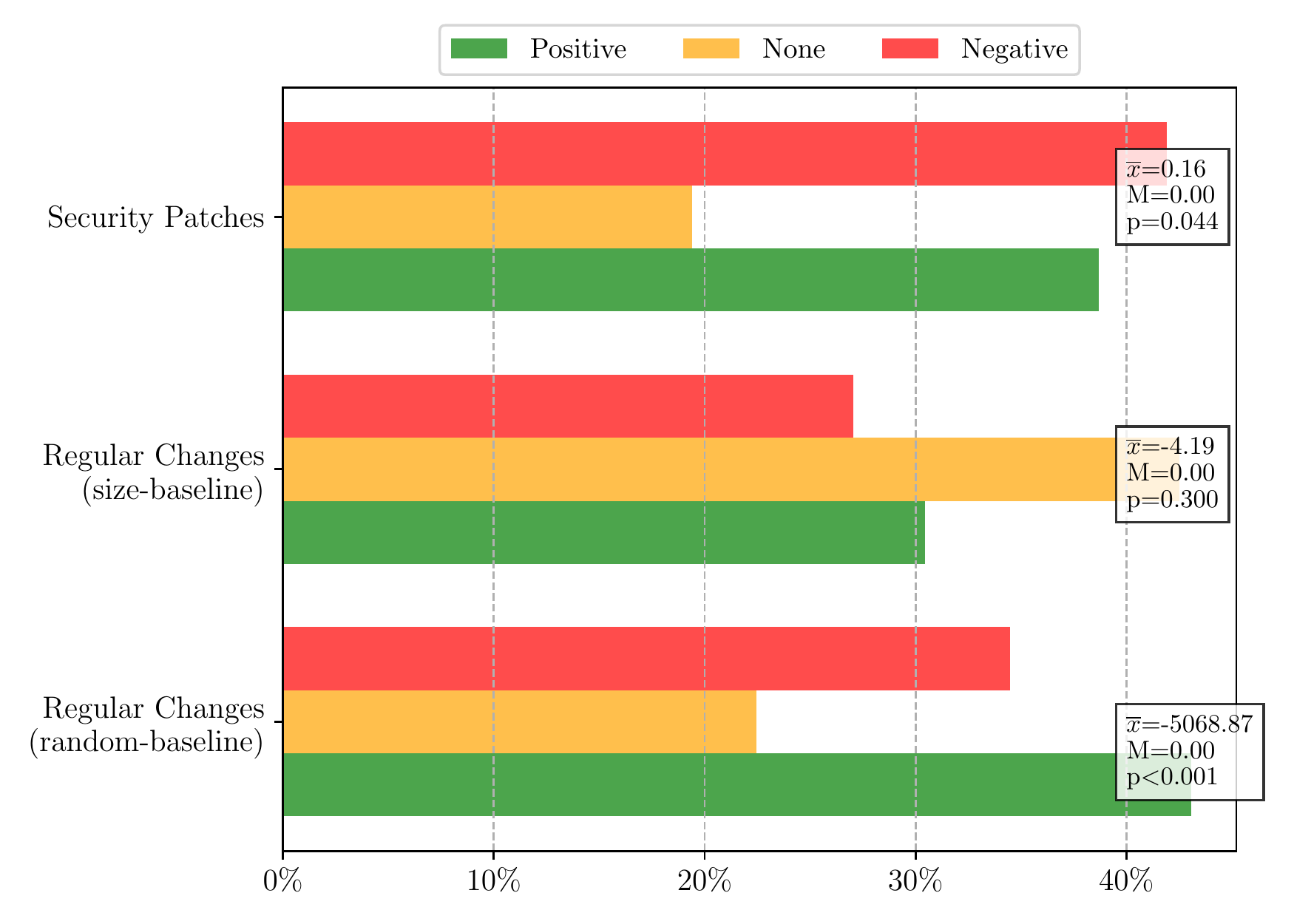}
  \caption{Maintainability difference of security patches versus regular changes.}
  \label{fig:secvsreg}    
\end{figure}

\textit{\textbf{RQ3: What is the impact of security patches versus regular 
changes on the maintainability of open-source software?}}
The impact of security and regular changes on software maintainability 
is presented in Figure~\ref{fig:secvsreg}. In this section, 
we present a comparison of security patches with 
two different baselines of regular changes: 
\emph{size-baseline}, a dataset of  
random regular changes with the same size as security 
patches---we argue that comparing changes with considerable different 
sizes may be unfair; and, \emph{random-baseline},
a dataset of random regular changes. 

Our hypothesis is
that \emph{security patches hinder more software maintainability 
than regular changes}. 
We have seen, previously, a deterioration in software maintainability 
when patching vulnerabilities: $41.9\%$ ($406$) of patches suffered a 
negative impact, $38.7\%$ ($375$) of patches remained the same, and $19.4\%$ 
($188$) of patches increased software maintainability. For regular changes,
when considering the size of the changes (\emph{size-basline}),
we observe that the maintainability decreases in $27.0\%$ ($262$) 
and increases in $30.5\%$ ($295$) of the cases. But in contrast to 
security patches, the maintainability of regular changes remains the 
same in $42.5\%$ ($412$) of the cases, i.e., performing regular
changes has a more positive impact than negative on  
maintainability. However, no statistical significance was 
obtained. Regular changes (\emph{random-baseline}), with no size 
restrictions, are less prone to hinder software maintainability than 
security changes. About $34.4\%$ ($333$) of the regular changes hinder 
software maintainability---less than in the security patches. For the \emph{random-baseline}, 
statistical significance was retrieved ($p = 5.34$x$10^{-8}$).

Overall, the results for both baselines show that regular 
changes are less prone to hinder the software 
maintainability of open-source software.   
However, the \emph{size-baseline} integrates a larger 
number of cases with no impact on software maintainability. 
We manually inspected a total of $25$ cases from
that distribution of regular changes with no impact on maintainability, 
and found that identifying regular changes with the same 
size as the security-related commit is limiting the type of regular 
commits being randomly chosen: input 
patches, variables or functions, type conversion (i.e., changes
with no impact on the software metrics analyzed by BCH).
We presume that this phenomenon lead
to the significant number of cases where there is no impact
on the software maintainability. On the other hand, 
identifying regular changes without any restrictions (\emph{random-baseline})
shows that regular changes have a less negative impact 
on software maintainability when compared to security patches
and that special attention should be given to security patches.

\textbf{Summary:} Security-related commits are observed to harm software 
maintainability, while regular changes are less prone to harm software maintainability.
Thus, we urge the importance of adopting maintainability practices while 
applying security patches.

\section{Study Implications}\label{sec:implications}
Our results show evidence that developers may have to reduce maintainability for the 
sake of security. We argue that developers should be able to patch and produce 
secure code without hindering the maintainability of their projects. But there are 
still concerns that need to be addressed and that this study brings awareness for:

\textit{\textbf{Follow Best Practices:}} Developers are not 
paying attention to some quality aspects of their solutions/patches, 
as seen in Figure~\ref{fig:guidelines}, ending up harming 
software maintainability. We argue that developers
should design and implement solutions that respect the 
limit bounds of branch points and function/module sizes that are 
recommended by best practices to avoid increasing the 
size, complexity, and dependencies of their patches.
Developers should also keep function parameters below 
the recommended limit. It helps keep unit interfaces small 
and easy to use and understand.
Patterns such as \emph{Introduce Parameter Object} are useful 
to send information to a new function/class through an object
and keep the number of parameters small and the information well-organized.

Security patches also harm the maintenance of software architecture. Maintaining the components independence and balance 
is important to make it easier to find the source code that 
developers want to patch/improve and to better understand
how the high-level components interact with others. Applying 
\emph{encapsulation} to hide implementation details and make the system 
more modular is a step forward not to hinder software architecture maintainability.

According to previous research, there is a correlation between 
code duplication and code smells~\cite{7476787}---duplicates are a source 
of regression bugs. BCH reports new code smells
for $15\%$ of the patches under study which supports previous research---$34\%$ of security patches 
introduce new problems~\cite{8819456,10.1145/3133956.3134072}. 
Developers should never reuse code by copying and pasting 
existing code fragments. Instead, they should create a method and call 
it every time needed. The \emph{Extract Method} refactoring 
technique solves many duplication problems. This makes spotting and 
solving the issue faster because you only need to fix one method instead
of multiple vulnerabilities. Clone detection tools such as CPD can help 
on locating duplicates.

\textit{\textbf{Prioritize High and Medium Severity:}} Previous research 
exhibits proof that developers prioritize
higher impact vulnerabilities~\cite{10.1145/3133956.3134072}.
Our study shows that vulnerabilities of high and
medium severity should be prioritized in software maintainability tasks.

\textit{\textbf{Some Types of Vulnerabilities Need More Attention:}} 
Our study attempted to shed light on the impact of different types of vulnerabilities on software maintainability.
Overall, all the CWEs under study present a negative impact
over $30\%$ on software maintainability. 
\emph{Cross-Site Scripting} (CWE-79) and \emph{Improper 
Restriction of Operations within the Bounds of a Memory Buffer} (CWE-119) 
are less prone to have an impact on open-source software maintainability.
Developers should pay special attention to \emph{Improper Input Validation} (CWE-20), 
\emph{Information Exposure} (CWE-200), \emph{Missing Release of Memory after Effective Lifetime} 
(CWE-401) and \emph{Path Traversal} (CWE-22). However, 
more research should be performed to better understand the impact of each guideline on each CWE.

\textit{\textbf{Tools for Patch Risk Assessment Wanted:}}
  Design debt of one guideline can lead to severe impacts on the 
  software quality~\cite{10.1145/1985362.1985366}. Some software 
  producers consider security as a first-class citizen while others do not. As 
  mentioned in previous work, security is critical and should be considered as a 
  default feature~\cite{10.1145/2489828.2489830,kurilova2014wyvern,mcgraw2004software}. 
  However, the lack of experts and awareness of developers for security while 
  producing/patching software leads companies to ship low-quality software. 
  Providing automated tools to developers to assess the risk of their patches is 
  essential to help companies shipping software of higher quality. 
  Bryan O'Sullivan, VP of Engineering at Facebook, advocated for new computer science
  risk models to detect vulnerabilities in scale and predict the level of security 
  of the software under production in his talk ```Challenges in Making Software Work at Scale''' at FaceTAV’20.

	Tools like Better Code Hub can complement static analysis (e.g., 
  SonarQube, Codacy, ESLint, Infer, and more) to provide more information 
  to security engineers on the vulnerabilities per se and the risk of their patches
  ---alongside pull requests/code reviews.
  Static code analysis may be daunting due 
	to the number of rules and different effects
	on maintainability. BCH claims to use the top-$10$ of the guidelines
	with the highest effects on maintainability. Yet, static analysis 
	can be used to complement the BCH analysis by introducing
	the capability of vulnerability detection and support the prediction 
	and prevention of the risk of a patch of hindering maintainability.

\textit{\textbf{Computer Science Curricula Needs to be Updated:}}
	Computer Science curricula is not yet prepared to properly educate 
	students for maintainable security. Students should be exposed to 
	this problem to gain experience with it. Curricula should focus on 
	the production of secure and maintainable code and alert to the 
	trade-off between both. These matters should be discussed and 
	presented to students in software engineering courses.
	Universities should also encourage students to use code quality 
	analysis tools such as BCH or similar ones. These tools have a 
	great potential to make students aware of maintainability issues 
	and beginner mistakes (e.g., coding practices violation).

\textit{\textbf{Secure Programming Languages by Design:}} 
In general, our study 
shows that there is over $35\%$ of patches with negative impact 
on software maintainability per programming language. Developers implementing code in   
\texttt{C/C++}, \texttt{Ruby} and \texttt{PHP} should pay extra care 
to their solutions/patches. In addition, programming 
	languages should provide new design patterns to easily patch security weaknesses 
	without endangering maintainability. Ultimately, new programming languages, 
	both secure and maintainable by design, such as Wyvern~\cite{10.1145/2489828.2489830,kurilova2014wyvern}, 
	should be designed to help developers be less vulnerability prone when 
	writing and maintaining secure applications. One example is the need for 
	designing new authorization mechanisms since these are one of the most 
	complex security features to implement.

\section{Threats to Validity}\label{sec:threats}
This section presents the potential threats to the validity of this study.

\textit{\textbf{Construct Validity:}} The formula to calculate the 
maintainability value ($M(v)$) was inferred based 
on the BCH's reports. The high amount of different projects and backgrounds 
may require other maintainability standards. However, BCH does use a 
representative benchmark of closed and open-source software projects to 
compute the thresholds for each maintainability guideline~\cite{Visser:2016:OREILLY,baggen2012}.
Maintainability is computed as the mean of all guidelines. Different software versions 
(vulnerable/fixed) of one vulnerability may have the same overall score
and still be affected by different guidelines. Therefore, we provide
an analysis per guideline, and our results are all available on GitHub for future reproductions and deeper analysis. 

\textit{\textbf{Internal Validity:}} The security patches dataset provided by previous
work~\cite{Reis:2017:IJSSE} was collected based on the messages of GitHub
commits produced by project developers to classify the changes performed while 
patching vulnerabilities. This approach discards patches that were
not explicit in commits messages. We assume that patches were performed using 
a single commit or several sequentially. The perspective that a 
developer may quickly perform a patch and later proceed to the refactor is 
not considered. We assume that all patches were only performed once. Depending 
on the impact of the vulnerability in the system, some vulnerabilities may have 
more urgency to be patched than others. For instance, a vulnerability performing a Denial-of-Service attack
that usually brings entire systems down may be more urgent to patch than a 
cross-site scripting vulnerability which generally does not have an impact
on the execution of the system but rather on the data accessibility.
We manually inspected $25.1\%$ of the security patches looking for floss-refactorings---$122$ from each dataset. We did 
find $23$ cases we argue to be floss-refactorings and toss them
to minimize the impact of this threat.

Baseline commits are retrieved randomly from 
the same project as the security patch.
This approach softens the differences
that may result from the characteristics of each project. However,
maintainability may still be affected by the developers' experience, coding
style, and software contribution policies which are not evaluated in this study.
Furthermore, this evaluation considers that $969$ regular commits---any kind
of commit---are enough to
alleviate random irregularities in the maintainability differences of the
baseline. 

\textit{\textbf{External Validity:}} The BCH tool uses private and open-source 
data to determine the thresholds for each guideline. We only analyze patches of open-source software.
Thus, our findings may not extend to private/non-open source software. 
Different programming languages may require different coding practices to address 
software safety. The dataset comprises more commits in Java, i.e., the dataset 
may not be representative of the population regarding programming languages. 
For both datasets, manual validation of the message of the commits was performed. 
Only commits in English were considered. Thus, our approach does not consider 
patches in any other language but English.

\section{Related Work}\label{sec:rw}

Many studies have investigated the relationship between patches and
software quality. Previous work focused on object-oriented metrics has evaluated the
impact of patches and exhibited proof that quantifying the
impact of patches on maintainability may help to choose the appropriate
patch type~\cite{1167822}. In contrast to this work, Hegedus et
al.~\cite{HEGEDUS2018313} did not select particular metrics to assess the effect
of patches. Instead, statistical tests were used to find the metrics that
have the potential to change significantly after patches. 

Researchers
performed a large-scale empirical study to understand the characteristics of security patches
and their differences against bug fixes~\cite{10.1145/3133956.3134072}.
The main findings were that security patches are smaller and less complex than bug 
fixes and are usually performed at the function level. Our study compares
the impact of security patches on software maintainability with the impact 
of regular changes.

Studying the evolution 
of maintainability issues during the development of Android apps, Malavolta et al. ($2018$)~\cite{8530041}
discovered that maintainability decreases over time. Palomba et al.
($2018$)~\cite{Palomba:2018:DIM:3231288.3231337} exhibits proof that code smells
should be carefully monitored by programmers since there is a high correlation
between maintainability aspects and proneness to changes/faults. In 2019, Cruz et 
al.~\cite{8919169} proposed a formula to calculate maintainability based on the 
BCH's guidelines and measured the impact of energy-oriented fixes 
on software maintainability. Recent work
proposed a new maintainability model to measure fine-grained code changes by 
adapting/extending the BCH model~\cite{8785997}.
Our work uses the same base model (SIG-MM) but considers a broader set of guidelines. 
Moreover, we solely focus on evaluating the impact of security patches on software maintainability.

Researchers investigated the relationship between design patterns and
maintainability~\cite{10.1007/978-3-642-35267-6-18}. However, other studies 
show that the use of design patterns may introduce maintainability issues into
software~\cite{4493325}. Yskout et. al did not detect if the usage of design 
patterns has a positive impact but concluded that developers prefer to work with 
the support of security patterns~\cite{8077802}. The present work studies how 
security weaknesses influence maintainability for open-source software.

There are studies that investigated the impact of programming languages on software
quality~\cite{Ray:2014:LSS:2635868.2635922,Ray:2017:LSP:3144574.3126905}. The first
one shows that some programming languages are more buggy-prone than others. However,
the authors of the second one could not reproduce it and did not obtain any
evidence about the language design impact. 
Berger et al. ($2019$)~\cite{2019arXiv190110220B} tried to reproduce~\cite{Ray:2014:LSS:2635868.2635922,Ray:2017:LSP:3144574.3126905} 
and identified flaws that throw into distrust the 
previously demonstrated a correlation between programming language and software 
defects. Our work studies how security patches affect software quality based 
on the code maintainability analysis and provides shows that programming languages 
may have an impact on maintainability.

\section{Conclusion and Future Work}\label{sec:conclusions}

This work presents an empirical study on the impact of $969$ security
patches on the maintainability of $260$ open-source projects. We leveraged
Better Code Hub reports to calculate maintainability based on a model proposed 
in previous work~\cite{Olivari:2018,8919169}. Results show evidence of a 
trade-off between security and maintainability, as $41.9\%$ of security patches 
yielded a negative impact. Hence,
developers may be hindering software maintainability while patching vulnerabilities. 
We also observe that some guidelines and programming languages are more likely to 
be affected than others. The implications of our study are that changes to codebases 
while patching vulnerabilities need to be performed with extra care; tools
for patch risk assessment should be integrated into the CI/CD pipeline; computer science
curricula need to be updated; and more secure programming languages are necessary.

As future work, the study can be extended in several directions: 
investigate which guidelines affect most the maintainability per
weakness; check if vulnerability patches are followed by new
commits and how much time does it take to do it; 
expand our methodology with other software quality properties; 
validate these findings with closed/private
software; and, expand this analysis to other quality standards.

\section*{Acknowledgements}

We thank SIG’s Better Code Hub team for all the support as
well as help in validating our methodology and results; and, Pedro Adão 
for the invaluable feedback in the early stages of the project.

This work is financed by National Funds through the
Portuguese funding agency, FCT - Fundação para a Ciência
e a Tecnologia with reference UIDB/50021/2020, a PhD scholarship 
(ref. SFRH/BD/143319/2019), the
SecurityAware Project (ref. CMU/TIC/0064/2019)---also funded
by the Carnegie Mellon Program---, and
the FaultLocker Project (ref. PTDC/CCI-COM/29300/2017).

\bibliographystyle{spmpsci}
\bibliography{emse20}   

\begin{thebibliography}{10}
\providecommand{\url}[1]{{#1}}
\providecommand{\urlprefix}{URL }
\expandafter\ifx\csname urlstyle\endcsname\relax
  \providecommand{\doi}[1]{DOI~\discretionary{}{}{}#1}\else
  \providecommand{\doi}{DOI~\discretionary{}{}{}\begingroup
  \urlstyle{rm}\Url}\fi

\bibitem{8077802}
{Acar}, Y., {Stransky}, C., {Wermke}, D., {Weir}, C., {Mazurek}, M.L., {Fahl},
  S.: Developers need support, too: A survey of security advice for software
  developers.
\newblock In: 2017 IEEE Cybersecurity Development (SecDev), pp. 22--26 (2017).
\newblock \doi{10.1109/SecDev.2017.17}

\bibitem{6113040}
{Alves}, T.L., {Correia}, J.P., {Visser}, J.: Benchmark-based aggregation of
  metrics to ratings.
\newblock In: 2011 Joint Conference of the 21st International Workshop on
  Software Measurement and the 6th International Conference on Software Process
  and Product Measurement, pp. 20--29 (2011).
\newblock \doi{10.1109/IWSM-MENSURA.2011.15}

\bibitem{5609747}
{Alves}, T.L., {Ypma}, C., {Visser}, J.: Deriving metric thresholds from
  benchmark data.
\newblock In: 2010 IEEE International Conference on Software Maintenance, pp.
  1--10 (2010).
\newblock \doi{10.1109/ICSM.2010.5609747}

\bibitem{baggen2012}
Baggen, R., Correia, J.P., Schill, K., Visser, J.: Standardized code quality
  benchmarking for improving software maintainability.
\newblock Software Quality Journal \textbf{20}(2), 287--307 (2012).
\newblock \doi{10.1007/s11219-011-9144-9}

\bibitem{2019arXiv190110220B}
{Berger}, E.D., {Hollenbeck}, C., {Maj}, P., {Vitek}, O., {Vitek}, J.: On the
  impact of programming languages on code quality.
\newblock arXiv e-prints arXiv:1901.10220 (2019)

\bibitem{Bijlsma:2012:FIR:2317098.2317124}
Bijlsma, D., Ferreira, M.A., Luijten, B., Visser, J.: Faster issue resolution
  with higher technical quality of software.
\newblock Software Quality Journal \textbf{20}(2), 265--285 (2012).
\newblock \doi{10.1007/s11219-011-9140-0}.
\newblock \urlprefix\url{http://dx.doi.org/10.1007/s11219-011-9140-0}

\bibitem{10.1145/1774088.1774504}
Chowdhury, I., Zulkernine, M.: Can complexity, coupling, and cohesion metrics
  be used as early indicators of vulnerabilities?
\newblock In: Proceedings of the 2010 ACM Symposium on Applied Computing, SAC
  '10, p. 1963–1969. Association for Computing Machinery, New York, NY, USA
  (2010).
\newblock \doi{10.1145/1774088.1774504}.
\newblock \urlprefix\url{https://doi.org/10.1145/1774088.1774504}

\bibitem{common:2009}
{Common Criteria Working Group}: Common methodology for information technology
  security evaluation.
\newblock Tech. rep., Technical report, Common Criteria Interpretation
  Management Board (2009)

\bibitem{8919169}
{Cruz}, L., {Abreu}, R., {Grundy}, J., {Li}, L., {Xia}, X.: Do energy-oriented
  changes hinder maintainability?
\newblock In: 2019 IEEE International Conference on Software Maintenance and
  Evolution (ICSME), pp. 29--40 (2019)

\bibitem{8785997}
{di Biase}, M., {Rastogi}, A., {Bruntink}, M., {van Deursen}, A.: The delta
  maintainability model: Measuring maintainability of fine-grained code
  changes.
\newblock In: 2019 IEEE/ACM International Conference on Technical Debt
  (TechDebt), pp. 113--122 (2019)

\bibitem{8819456}
{Elkhail}, A.A., {Cerny}, T.: On relating code smells to security
  vulnerabilities.
\newblock In: 2019 IEEE 5th Intl Conference on Big Data Security on Cloud
  (BigDataSecurity), IEEE Intl Conference on High Performance and Smart
  Computing, (HPSC) and IEEE Intl Conference on Intelligent Data and Security
  (IDS), pp. 7--12 (2019)

\bibitem{oswap:2013}
Foundation, T.O.: Owasp top 10 - 2017: The ten most critical web application
  security risks.
\newblock Tech. rep., The OWASP Foundation (2017).
\newblock Release Candidate

\bibitem{oswap:2017}
Foundation, T.O.: Owasp top 10 - 2017: The ten most critical web application
  security risks.
\newblock Tech. rep., The OWASP Foundation (2017).
\newblock Release Candidate

\bibitem{10.1007/978-3-642-35267-6-18}
Heged{\H{u}}s, P., B{\'a}n, D., Ferenc, R., Gyim{\'o}thy, T.: Myth or reality?
  analyzing the effect of design patterns on software maintainability.
\newblock In: Computer Applications for Software Engineering, Disaster
  Recovery, and Business Continuity, pp. 138--145. Springer Berlin Heidelberg,
  Berlin, Heidelberg (2012)

\bibitem{HEGEDUS2018313}
Hegedűs, P., Kádár, I., Ferenc, R., Gyimóthy, T.: Empirical evaluation of
  software maintainability based on a manually validated refactoring dataset.
\newblock Information and Software Technology \textbf{95}, 313 -- 327 (2018).
\newblock \doi{https://doi.org/10.1016/j.infsof.2017.11.012}

\bibitem{4335232}
{Heitlager}, I., {Kuipers}, T., {Visser}, J.: A practical model for measuring
  maintainability.
\newblock In: 6th International Conference on the Quality of Information and
  Communications Technology (QUATIC 2007), pp. 30--39 (2007).
\newblock \doi{10.1109/QUATIC.2007.8}

\bibitem{iso:2011}
{International Organization for Standardization}: International Standard
  ISO/IEC 25010. Systems and Software Engineering - Systems and software
  Quality Requirements and Evaluation (SQuaRE) - System and Software Quality
  Models (2011)

\bibitem{7476787}
{Islam}, M.R., {Zibran}, M.F.: A comparative study on vulnerabilities in
  categories of clones and non-cloned code.
\newblock In: 2016 IEEE 23rd International Conference on Software Analysis,
  Evolution, and Reengineering (SANER), vol.~3, pp. 8--14 (2016)

\bibitem{just2014mutants}
Just, R., Jalali, D., Inozemtseva, L., Ernst, M.D., Holmes, R., Fraser, G.: Are
  mutants a valid substitute for real faults in software testing?
\newblock In: Proceedings of the 22nd ACM SIGSOFT International Symposium on
  Foundations of Software Engineering, pp. 654--665. ACM (2014)

\bibitem{1167822}
Kataoka, Y., Imai, T., Andou, H., Fukaya, T.: A quantitative evaluation of
  maintainability enhancement by refactoring.
\newblock In: International Conference on Software Maintenance, 2002.
  Proceedings., pp. 576--585 (2002).
\newblock \doi{10.1109/ICSM.2002.1167822}

\bibitem{4493325}
Khomh, F., Gueheneuce, Y.: Do design patterns impact software quality
  positively?
\newblock In: 2008 12th European Conference on Software Maintenance and
  Reengineering, pp. 274--278 (2008).
\newblock \doi{10.1109/CSMR.2008.4493325}

\bibitem{kurilova2014wyvern}
Kurilova, D., Potanin, A., Aldrich, J.: Wyvern: Impacting software security via
  programming language design.
\newblock In: Proceedings of the 5th Workshop on Evaluation and Usability of
  Programming Languages and Tools, pp. 57--58 (2014)

\bibitem{10.1145/3133956.3134072}
Li, F., Paxson, V.: A large-scale empirical study of security patches.
\newblock In: Proceedings of the 2017 ACM SIGSAC Conference on Computer and
  Communications Security, CCS '17, p. 2201–2215. Association for Computing
  Machinery, New York, NY, USA (2017).
\newblock \doi{10.1145/3133956.3134072}.
\newblock \urlprefix\url{https://doi.org/10.1145/3133956.3134072}

\bibitem{8530041}
Malavolta, I., Verdecchia, R., Filipovic, B., Bruntink, M., Lago, P.: How
  maintainability issues of android apps evolve.
\newblock In: 2018 IEEE International Conference on Software Maintenance and
  Evolution (ICSME), pp. 334--344 (2018).
\newblock \doi{10.1109/ICSME.2018.00042}

\bibitem{4724577}
Maruyama, K., Tokoda, K.: Security-aware refactoring alerting its impact on
  code vulnerabilities.
\newblock In: 2008 15th Asia-Pacific Software Engineering Conference, pp.
  445--452 (2008).
\newblock \doi{10.1109/APSEC.2008.57}

\bibitem{1702388}
McCabe, T.J.: A complexity measure.
\newblock IEEE Transactions on Software Engineering \textbf{SE-2}(4), 308--320
  (1976).
\newblock \doi{10.1109/TSE.1976.233837}

\bibitem{mcgraw2004software}
McGraw, G.: Software security.
\newblock IEEE Security \& Privacy \textbf{2}(2), 80--83 (2004)

\bibitem{graw:1992}
McGraw, K.O., Wong, S.P.: A common language effect size statistic.
  psychological bulletin  (1992).
\newblock \doi{doi:10.1037/0033-2909.111.2.361}

\bibitem{10.1145/2489828.2489830}
Nistor, L., Kurilova, D., Balzer, S., Chung, B., Potanin, A., Aldrich, J.:
  Wyvern: A simple, typed, and pure object-oriented language.
\newblock In: Proceedings of the 5th Workshop on MechAnisms for SPEcialization,
  Generalization and InHerItance, MASPEGHI '13, p. 9–16. Association for
  Computing Machinery, New York, NY, USA (2013).
\newblock \doi{10.1145/2489828.2489830}.
\newblock \urlprefix\url{https://doi.org/10.1145/2489828.2489830}

\bibitem{Olivari:2018}
Olivari, M.: {Maintainable Production: A Model of Developer Productivity Based
  on Source Code Contributions}.
\newblock Master's thesis, University of Amsterdam (2018)

\bibitem{Palomba:2018:DIM:3231288.3231337}
Palomba, F., Bavota, G., Penta, M.D., Fasano, F., Oliveto, R., Lucia, A.D.: On
  the diffuseness and the impact on maintainability of code smells: A large
  scale empirical investigation.
\newblock Empirical Softw. Engg. \textbf{23}(3), 1188--1221 (2018).
\newblock \doi{10.1007/s10664-017-9535-z}

\bibitem{10.1109/MSR.2019.00064}
Ponta, S.E., Plate, H., Sabetta, A., Bezzi, M., Dangremont, C.: A
  manually-curated dataset of fixes to vulnerabilities of open-source software.
\newblock In: Proceedings of the 16th International Conference on Mining
  Software Repositories, MSR ’19, p. 383–387. IEEE Press (2019).
\newblock \doi{10.1109/MSR.2019.00064}.
\newblock \urlprefix\url{https://doi.org/10.1109/MSR.2019.00064}

\bibitem{Pothamsetty:2005:SEL:1107622.1107635}
Pothamsetty, V.: Where security education is lacking.
\newblock In: Proceedings of the 2Nd Annual Conference on Information Security
  Curriculum Development, InfoSecCD '05, pp. 54--58. ACM, New York, NY, USA
  (2005).
\newblock \doi{10.1145/1107622.1107635}.
\newblock \urlprefix\url{http://doi.acm.org/10.1145/1107622.1107635}

\bibitem{10.2307/2282543}
Pratt, J.W.: Remarks on zeros and ties in the wilcoxon signed rank procedures.
\newblock Journal of the American Statistical Association \textbf{54}(287),
  655--667 (1959)

\bibitem{Ray:2017:LSP:3144574.3126905}
Ray, B., Posnett, D., Devanbu, P., Filkov, V.: A large-scale study of
  programming languages and code quality in github.
\newblock Commun. ACM \textbf{60}(10), 91--100 (2017).
\newblock \doi{10.1145/3126905}

\bibitem{Ray:2014:LSS:2635868.2635922}
Ray, B., Posnett, D., Filkov, V., Devanbu, P.: A large scale study of
  programming languages and code quality in {Github}.
\newblock In: Proceedings of the 22Nd ACM SIGSOFT International Symposium on
  Foundations of Software Engineering, FSE 2014, pp. 155--165. ACM, New York,
  NY, USA (2014).
\newblock \doi{10.1145/2635868.2635922}

\bibitem{Reis:2017:IJSSE}
Reis, S., Abreu, R.: A database of existing vulnerabilities to enable
  controlled testing studies.
\newblock International Journal of Secure Software Engineering (IJSSE)
  \textbf{8}(3) (2017).
\newblock \doi{10.4018/IJSSE.2017070101}

\bibitem{reis2017secbench}
Reis, S., Abreu, R.: Secbench: A database of real security vulnerabilities.
\newblock In: Proceedings of the International Workshop on Secure Software
  Engineering in DevOps and Agile Development (SecSE 2017) (2017)

\bibitem{schneier2006beyond}
Schneier, B.: Beyond fear: Thinking sensibly about security in an uncertain
  world.
\newblock Springer Science \& Business Media (2006)

\bibitem{shin2010evaluating}
Shin, Y., Meneely, A., Williams, L., Osborne, J.A.: Evaluating complexity, code
  churn, and developer activity metrics as indicators of software
  vulnerabilities.
\newblock IEEE transactions on software engineering \textbf{37}(6), 772--787
  (2010)

\bibitem{slaughter1998evaluating}
Slaughter, S.A., Harter, D.E., Krishnan, M.S.: Evaluating the cost of software
  quality.
\newblock Communications of the ACM \textbf{41}(8), 67--73 (1998)

\bibitem{4267025}
{Telang}, R., {Wattal}, S.: An empirical analysis of the impact of software
  vulnerability announcements on firm stock price.
\newblock IEEE Transactions on Software Engineering \textbf{33}(8), 544--557
  (2007)

\bibitem{oswap:2009}
{The OWASP Foundation}: {OWASP} application security verification standard 2009
  - web application standard.
\newblock Tech. rep. (2009)

\bibitem{Visser:2016:OREILLY}
Visser, J.: Building Maintainable Software, Java Edition: Ten Guidelines for
  Future-Proof Code.
\newblock O'Reilly Media, Inc.

\bibitem{criteria:2017}
Visser, J.: Sig/tÜvit evaluation criteria trusted product maintainability:
  Guidance for producers.
\newblock Available: \url{https://bit.ly/2H8QZBo} (2018)

\bibitem{10.2307/3001968}
Wilcoxon, F.: Individual comparisons by ranking methods.
\newblock Biometrics Bulletin \textbf{1}(6), 80--83 (1945)

\bibitem{6616351}
Xu, H., Heijmans, J., Visser, J.: A practical model for rating software
  security.
\newblock In: 2013 IEEE Seventh International Conference on Software Security
  and Reliability Companion, pp. 231--232 (2013).
\newblock \doi{10.1109/SERE-C.2013.11}

\bibitem{10.1145/1985362.1985366}
Zazworka, N., Shaw, M.A., Shull, F., Seaman, C.: Investigating the impact of
  design debt on software quality.
\newblock In: Proceedings of the 2nd Workshop on Managing Technical Debt, MTD
  ’11, p. 17–23. Association for Computing Machinery, New York, NY, USA
  (2011).
\newblock \doi{10.1145/1985362.1985366}.
\newblock \urlprefix\url{https://doi.org/10.1145/1985362.1985366}

\bibitem{5773403}
{Zibran}, M.F., {Saha}, R.K., {Asaduzzaman}, M., {Roy}, C.K.: Analyzing and
  forecasting near-miss clones in evolving software: An empirical study.
\newblock In: 2011 16th IEEE International Conference on Engineering of Complex
  Computer Systems, pp. 295--304 (2011)

\end{thebibliography}

\end{document}